%% file: THz_Magazine.tex
\documentclass[10pt,conference]{IEEEtran}
\usepackage{setspace}
\usepackage{amsmath}
\usepackage{acronym}
\usepackage{bbm}
\usepackage[dvips]{color}
\usepackage{comment}
\usepackage{todonotes}
\usepackage{epsf}
\usepackage{siunitx}
\usepackage{epsfig}
\usepackage{times}
\usepackage{epsfig}
\usepackage{graphicx}
\usepackage{paralist}
\usepackage{bbold}
\usepackage{enumitem}
\usepackage{geometry}
\usepackage{mathrsfs}
\usepackage{amssymb}
\usepackage{pdfpages}
\usepackage{epstopdf}
\usepackage{tcolorbox,bbold}
\usepackage{algorithm,algorithmicx,algpseudocode}
\makeatletter
\newcommand{\algmargin}{\the\ALG@thistlm}
\makeatother
\algnewcommand{\parState}[1]{\State%
	\parbox[t]{\dimexpr\linewidth-\algmargin}{\strut #1\strut}}

\usepackage[nomain,acronym,shortcuts]{glossaries}
\newcommand*{\acro}[3][]{\newacronym[#1]{#2}{#2}{#3}}
\include{acronyms}

\usepackage{amssymb}
\usepackage{url}
\usepackage{dsfont}
\usepackage{lettrine} 
\usepackage{amsmath,epsfig,amssymb,algorithm,algpseudocode,amsthm,cite,url}
\IEEEoverridecommandlockouts
\usepackage{subcaption}
\allowdisplaybreaks
\usepackage{csquotes}

\usepackage{verbatim}
\usepackage[english]{babel}
\usepackage{amsmath,amssymb}

\usepackage[justification=centering]{caption}
\usepackage{verbatim}

\IEEEoverridecommandlockouts
\begin{document}
	\title{\huge{Seven Defining Features of Terahertz (THz) Wireless Systems: A Fellowship of Communication and Sensing}}
		\author{Christina Chaccour, \emph{Student Member, IEEE,}
			Mehdi Naderi Soorki, Walid Saad, \emph{Fellow, IEEE,}\\ Mehdi Bennis, \emph{Fellow, IEEE,} Petar Popovski, \emph{Fellow, IEEE}, and  M\'erouane Debbah,  \emph{Fellow, IEEE}
		\thanks{\noindent This research was supported by the U.S. National Science Foundation under Grants CNS-1836802 and AST-2037870.}
		\thanks{C. Chaccour and W. Saad are with the Wireless@VT, Bradley Department of Electrical and Computer Engineering, Virginia Tech, Blacksburg, VA, USA, Emails: \protect{christinac@vt.edu}, \protect{walids@vt.edu}.}
		\thanks{M. Naderi Soorki is with the Department of Electrical and Computer
			Engineering, Isfahan University of Technology, Isfahan 84156-83111, Email: \protect{mehdin@vt.edu}.}
		\thanks{M. Bennis is with the Centre for Wireless Communications, University
			of Oulu, 90014 Oulu, Finland, Email: \protect{mehdi.bennis@oulu.fi}.}
		\thanks{P. Popovski is with the Department of Electronic Systems, Aalborg University, Denmark, Email: \protect{petarp@es.aau.dk}.}
	\thanks{ M\'erouane Debbah is with {Université Paris-Saclay, CNRS, CentraleSupélec, Gif-sur-Yvette, France and the Lagrange Mathematical and Computing Research Center, Paris, France, Email: \protect{merouane.debbah@centralesupelec.fr}}}}
	\maketitle
	%
	

	\begin{abstract}
Wireless communication at the terahertz (THz) frequency bands ($0.1-\SI{10}{THz}$) is viewed as one of the cornerstones of tomorrow's 6G wireless systems. Owing to the large amount of available bandwidth, if properly deployed, THz frequencies can potentially provide significant wireless capacity performance gains and enable high-resolution environment sensing. However, operating a wireless system at high-frequency bands such as THz is limited by a highly uncertain and dynamic channel. Effectively, these channel limitations lead to unreliable intermittent links as a result of an inherently short communication range, and a high susceptibility to blockage and molecular absorption. Consequently, such impediments could disrupt the THz band's promise of \emph{high-rate communications and high-resolution sensing} capabilities. In this context, this paper panoramically examines the steps needed to efficiently and reliably deploy and operate next-generation THz wireless systems that will synergistically support a fellowship of communication and sensing services. For this purpose, we first set the stage by describing the fundamentals of the THz frequency band. Based on these fundamentals, we characterize and comprehensively investigate \emph{seven unique defining features of THz wireless systems}: 1) Quasi-opticality of the band, 2) THz-tailored wireless architectures, 3) Synergy with lower frequency bands, 4) Joint sensing and communication systems, 5) PHY-layer procedures, 6) Spectrum access techniques, and 7) Real-time network optimization. These seven defining features allow us to shed light on how to re-engineer wireless systems as we know them today so as to make them ready to support THz bands and their unique environments. On the one hand, THz systems benefit from their quasi-opticality and \emph{can turn every communication challenge into a sensing opportunity}, thus contributing to a new generation of  \emph{versatile wireless systems} that can perform multiple functions beyond basic communications. On the other hand, THz systems can capitalize on the role of intelligent surfaces, lower frequency bands, and machine learning (ML) tools to guarantee a robust system performance. We conclude our exposition by presenting the key THz 6G use cases along with their associated major challenges and open problems. Ultimately, the goal of this article is to chart a forward-looking roadmap that exposes the necessary solutions and milestones for enabling THz frequencies to realize their potential as a game changer for next-generation wireless systems.\\
	{ \indent \emph{Index Terms}--- Terahertz (THz); 6G; Internet of Everything (IoE); Sensing; Wireless Systems; Joint Sensing and Communication Systems; Machine Learning (ML).}
\end{abstract}
\section{Introduction}
The sixth generation (6G) of wireless cellular systems must cater to radically new services, such as immersive remote presence, holographic teleportation, \ac{CRAS}, \ac{XR}, and digital twins. These bandwidth-intensive applications require the delivery of $1000\times$ capacity increase \cite{saad2019vision} compared to what is expected from today's 5G cellular systems. These applications also require multi-purpose wireless functions that could encompass communications, sensing, localization, and control. These requirements can only be attained by boosting the existing wireless spectrum bands at sub-6 GHz and \ac{mmWave} with abundant bandwidth through a migration towards the higher frequency \ac{THz} bands \cite{sarieddeen2020overview}. In particular, the \ac{THz} band, namely $0.1-\SI{10}{THz}$\footnote{The terahertz gap consists of the band $0.1-\SI{10}{THz}$, whereas the frequencies in the range $0.1-\SI{0.3}{THz}$ are typically considered to be sub-THz band. According to the ITU \cite{ITUreport}, the band above 275 GHz exhibits the unique THz properties and is the main part of the THz band. Nonetheless, this article targets the continuum of the sub-THz and THz bands, as a result of the prevailing feasibility of transceiver design at the sub-THz band. For simplicity of nomenclature, we will use the term \ac{THz} to refer to this entire range.}, was the last gap in the spectrum to be bridged and provides a golden mean between radio and optical signals. As such, the late discovery of the \ac{THz} gap and the unknown peculiarities of its behavior delayed the deployment of the \ac{THz} frequency band in real-world wireless networks, however, this status is rapidly changing today \cite{elayan2019terahertz}.\\
	\begin{figure*} [t!]
	\begin{centering}
		\includegraphics[width=.8\textwidth]{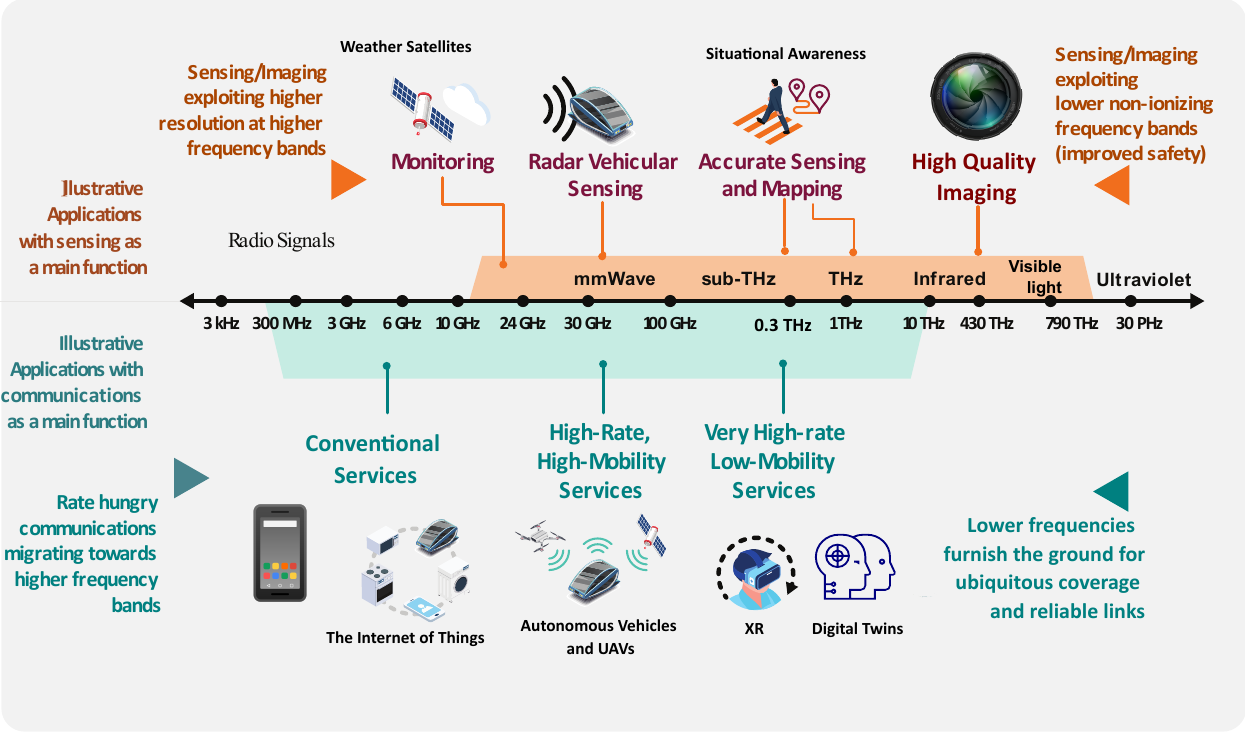}
		\caption{\small{Illustrative figure showcasing the paradigm shift on the frequency spectrum whereby the sub-$\SI{6}{GHz}$ $-$ \ac{THz} region is being jointly populated by communication and sensing functionalities.}}
		\label{fig:frequency_spec]}
	\end{centering}
\end{figure*}
\indent The \emph{THz gap} results from the inability of producing efficient transceivers and antennas at \ac{THz} frequencies. In particular, at \ac{THz} bands, semiconductor devices fail to effectively convert electrical energy into electromagnetic energy. For example, at these frequencies, electrons cannot travel the distance needed to enable a semiconductor device to work, before the polarity of the potential voltage and electrons directions changes. Thus, on the one hand, these challenges at the transceiver level delayed real-world access to the \ac{THz} frequencies. On the other hand, prior to the emergence of the aforementioned 6G services, \ac{THz} frequencies exhibited more challenges than opportunities given their transceiver design difficulties and their cumbersome channel behavior.\\
\indent Owing to the recent advances in plasmonic devices and graphene-based designs, gradually, the aforementioned challenges at the transceiver level are being successfully mitigated \cite{leuthold2015plasmonic}. In contrast to conventional electronic and photonic-based designs, plasmonics do not rely on electrons or photons but, instead, they require electromagnetic excitation at the metal surface level at optical frequencies\cite{elayan2019terahertz}. Similarly, graphene-based designs provide interesting electrical and optical properties that allow transceivers to be tunable, compact, high-speed, and energy efficient. As these innovations started seeing light, \ac{THz} first gained popularity in the realm of nano-networks \cite{lemic2019assessing}. Transceiver design in this domain was less challenging given that the generation of a hundred femtosecond pulses at low power foreshadows the feasibility of short-range communication and nano-networks \cite{jornet2014femtosecond}. Connecting nano-machines, nano-sensors, and nano-actuators allows these devices to interact at the same scale of living systems and chipsets, ultimately resulting in the \ac{IoNT}. Thus, the rise of the \ac{IoNT} would potentially yield wireless innovations ranging from interconnecting chip networks, all the way to biosensors monitoring human and organ activity at the \ac{THz} band. Next, we discuss the specific technical enhancements needed to guarantee a successful \ac{THz} deployment for the \ac{IoE}. 
\subsection{Towards \ac{THz} for 6G and the \ac{IoE}}
\indent Recent evolutions at the transceiver design level are ushering in a transition of \ac{THz} communication from its limited \ac{IoNT} application domain to the realms of the \ac{IoE} and the forthcoming wireless 6G systems. Nonetheless, such a transition towards the real-world deployment of \ac{THz} as a means to provide communication and sensing services for \ac{IoE} applications faces many modeling, network analysis, design, and optimization challenges. \emph{First}, the characteristics of \ac{THz} frequencies require an evolution in the wireless system architecture so as to account for the highly varying channel, the short range of nature of the \ac{THz} links, and the dependence on narrowbeam \ac{LoS} links. \emph{Second}, well-designed \ac{THz} architectures, consisting of distributed and heterogeneous \acp{SBS}, necessitate novel approaches to estimate the \ac{THz} channel, maximize the \ac{THz} network coverage, and synchronize the \ac{THz} system resources with other frequency bands. \emph{Third}, emerging 6G services have considerably stringent reliability, latency, and rate requirements. In turn, meeting these requirements mandates novel real-time sensing, optimization, and scheduling approaches. As a result, a successful transition towards \ac{IoE} \ac{THz} deployment requires a significant rethinking of conventional physical (PHY) layer and networking procedures.\\
\indent In addition to delivering extremely high communication data rates, \emph{the quasi-optical nature of \ac{THz} frequency bands holds under its belt promising capabilities for sensing, imaging, and localization functions} that are not yet very well understood. On these grounds, it is worth noting that as shown in Fig.~\ref{fig:frequency_spec]}, the frequency spectrum is being jointly populated by communication and sensing functionalities in the sub-$\SI{6}{GHz} -$ \ac{THz} region. Effectively, this particular region has become attractive to wireless sensing and communication services as a result of several factors. \emph{First}, in contrast to the customary frequency bands adopted for imaging (e.g. X-Rays), this region has non-ionizing radiation. \emph{Second}, in contrast to optical links, the sensing \ac{EM} links have a less intermittent behavior. \emph{Third}, from a communication standpoint, this region offers various data rates ranging from average to extremely high data rates. As such, based on the properties of this populated region, the lower band (key players here are sub- $\SI{6}{GHz}$ and sub-\ac{mmWave} frequency bands) offers average data rates and sensing capabilities with longer ranges, higher reliability, and lower resolution. Meanwhile, and of rising interest, the higher band of this region (sub-\ac{THz} and \ac{THz} band) provides high-data rates and high-resolution sensing but at the expense of shorter distances and more intermittent links as reflected in Fig.~\ref{fig:frequency_spec]}. Furthermore, deploying the majority of services in the higher band of this region can help alleviate the challenge of spectrum scarcity for wireless communications that is experienced in the sub-6 GHz bands. Thus, \emph{given the particular benefits that can be procured in communications, sensing, and imaging at the \ac{THz} band, jointly deploying such systems allows sharing resources and provides \ac{THz} communications high-resolution localization and situational awareness.} Such configurations not only provide a higher spectrum efficiency, but also pave the way for novel opportunities stemming from the possibility of performing coordinated sensing and communications.
	\begin{figure*} [t!]
	\begin{centering}
		\includegraphics[width=.85\textwidth]{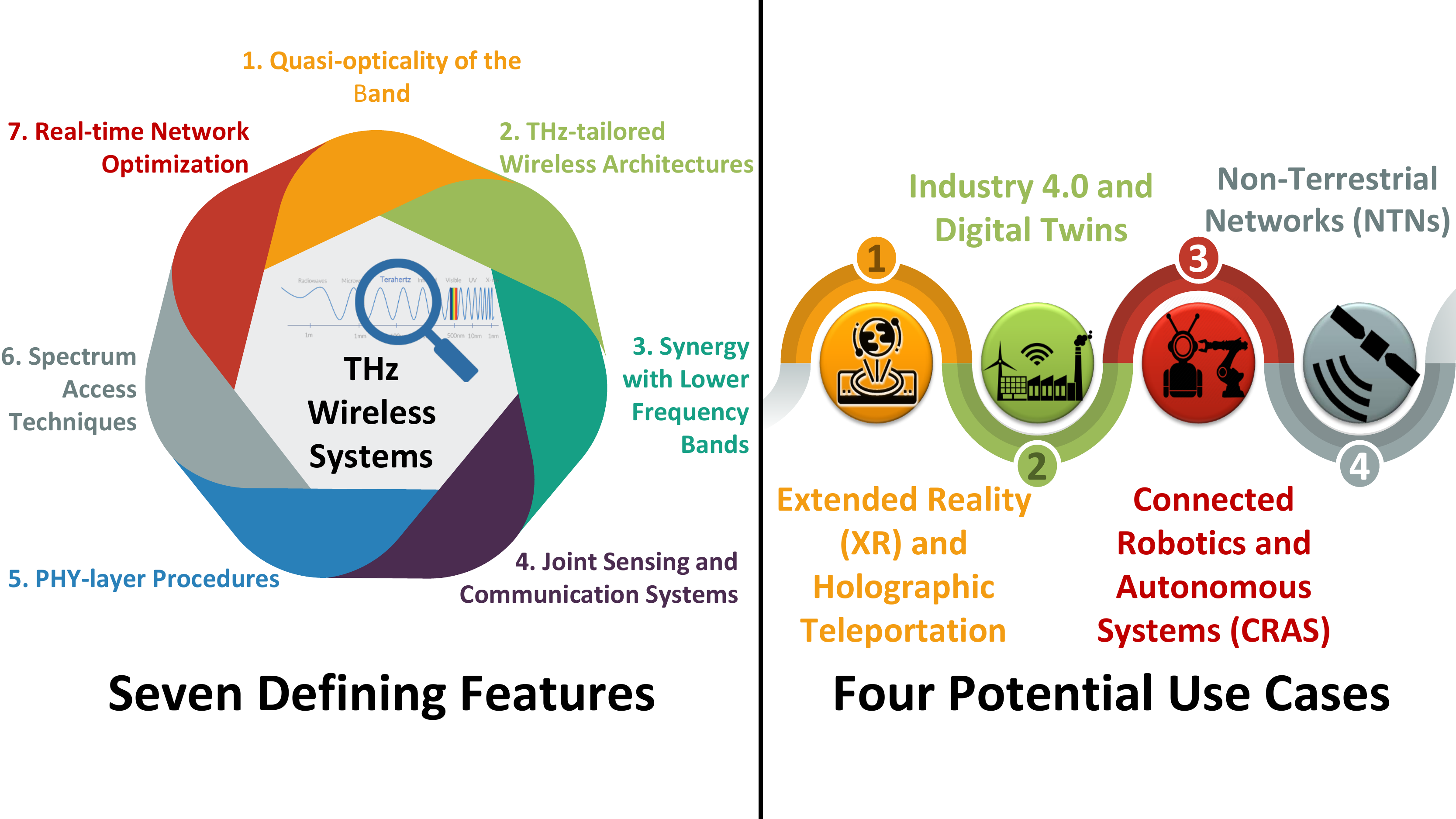}
		\caption{\small{Illustrative figure showcasing the seven unique defining features of wireless THz systems and their key use cases.}}
		\label{fig:comprehensive}
	\end{centering}
\end{figure*}
\subsection{Prior Works}
\indent Recently, a number of surveys and tutorials related to the deployment of \ac{THz} over wireless networks appeared in \cite{sarieddeen2020overview, elayan2019terahertz}, and \cite{petrov2018last, zhang2020beyond, peng2020channel, polese2020toward, sarieddeen2020next, ghafoor2020mac}. The works in \cite{petrov2018last, zhang2020beyond, peng2020channel} primarily focused on the PHY layer or propagation challenges. In fact, while such works discuss key channel models along with the caveats of \ac{THz} propagation, they do not scrutinize the quasi-optical properties of the \ac{THz} frequency band and its significance on the PHY layer challenges and opportunities at \ac{THz}.  Meanwhile, the works in  \cite{sarieddeen2020overview, elayan2019terahertz}, and \cite{polese2020toward, sarieddeen2020next, ghafoor2020mac} summarize the latest literature findings associated with the deployment of \ac{THz} wireless networks. Although this important prior art exposes key concepts about \ac{THz} networks, it does not propose transformative solutions that are needed to deploy realistic \ac{THz} networks effectively. Furthermore, these works do not articulate the challenges of adopting known channel estimation techniques and network optimization methods when dealing with the high uncertainty of the \ac{THz} channel. As such, they do not outline the breakthroughs required to defy the non-stationary and time varying \ac{THz} channel so as to provide initial access, maximize the \ac{THz} coverage, and enable a seamless network optimization. Moreover, although some of these works (e.g. \cite{sarieddeen2020overview} and \cite{sarieddeen2020next}) recognize the importance of the \ac{THz} sensing capability, they fail to highlight the challenges, opportunities, and the prospective techniques needed to deploy fully-fledged joint sensing and communication systems at \ac{THz} frequencies. 
\subsection{Contributions}
The main contribution of this article is a novel and holistic vision that articulates the unique role of \ac{THz} in next-generation wireless systems. We first examine the fundamentals of \ac{THz} frequency bands and their propagation characteristics. By leveraging this fundamental examination of the \ac{THz} properties, we identify and provide a comprehensive treatment of the \emph{seven unique features} that will define future \ac{THz} wireless systems. Subsequently, we examine the behavior and needs of each feature, and we propose opportunistic techniques that harvest peculiar \ac{THz} benefits. Hence, these benefits maximize the overall system performance and could potentially elevate  \ac{THz} wireless systems to another level. The seven unique characteristics that we envision to be the defining features of  future THz-based wireless systems are shown on the left-hand of Fig.~\ref{fig:comprehensive} and discussed next: 
\subsubsection{Quasi-opticality of the \ac{THz} band} The \ac{EM} properties of the \ac{THz} band, by virtue of its quasi-opticality lead to distinct communication challenges. Most importantly, the molecular absorption effect is seen as a limiting factor to the propagation of \ac{THz} waves. Nonetheless, we will underscore how this quasi-opticality is a double-edged sword for the foreseen joint communication and sensing paradigm. For instance, the molecular absorption opens up various sensing opportunities that are not found at other frequency bands. However, this comes at the expense of shorter communication ranges. We will also highlight the other sensing functionalities offered by this quasi-opticality.
\subsubsection{\ac{THz}-tailored architectures} Deploying wireless \ac{THz} systems requires accounting for a higher density of \acp{SBS}, a shorter communication range, the ability to deliver multiple functions (sensing, communication, imaging), and a set of unique channel conditions. These factors urge adopting more opportunistic \ac{THz}-tailored network architectures that can exploit the advantages of \ac{THz} systems. As such, we particularly emphasize the importance of adopting cell less architectures, as well as their accompanying challenges and opportunities. Furthermore, we highlight the pivotal role of \acp{RIS} in \ac{THz} networks \cite{chaccour2020risk, huang2020holographic, jung2020performance}, their holographic capability due to the small \ac{THz} footprint, their massive sensing elements, and the near-field communication opportunities and challenges.
\subsubsection{Synergy of \ac{THz} with lower frequency bands} Communication systems at \ac{THz} frequencies will be deployed in a radio spectrum that is already highly populated with sub-$\SI{6}{GHz}$ and \ac{mmWave} technologies. In that sense, \ac{THz} systems are expected to have a certain level of synergy (cooperation and seamless coexistence) with the lower frequency band wireless technologies. For instance, certain use cases, such as immersive remote presence, could opportunistically use all available wireless frequencies to deliver the target end-to-end experience. Thus, we underline the strategies that enable the coexistence of \ac{THz} frequencies with \ac{mmWave} and sub-$\SI{6}{GHz}$ bands, services, and infrastructure. We further point out how this synergy across the different frequency bands opens the door for exciting opportunities for both communication and sensing functionalities. 
\subsubsection{Joint sensing and communication systems} Owing to the quasi-optical nature of \ac{THz} bands, a harmonious fellowship of \emph{high-rate communications and high-resolution sensing} can be formed. Consequently, we accentuate the role of joint communication and sensing systems for future \ac{THz} wireless networks. Particularly, we emphasize the effectiveness of mutual feedback between the sensing and communication functionalities that can improve the overall system performance. Naturally, adopting such configurations can help transform wireless networks into a new generation of \emph{versatile} systems that can offer multiple functions to their users thus opening the door for novel services and use cases to be used at the \ac{THz} band.
\subsubsection{PHY-layer procedures} The spatially-sparse and low rank \ac{THz} channel imposes distinct challenges on the PHY-layer procedures such as wireless channel estimation and initial access. To overcome these challenges, we propose novel channel estimation techniques that bring to light the role of generative learning networks in predicting the full \ac{THz} \ac{CSI}. Furthermore, we highlight the role of sensing in ensuring an enhanced initial network access for \ac{THz} devices. 
\subsubsection{Spectrum access techniques} Conventional access schemes adopted in previous wireless generations cannot be directly applied to \ac{THz} frequency bands due to hardware constraints and the unique nature of the \ac{THz} propagation environment. Consequently, we examine possible spectrum access techniques that are suitable for \ac{THz} systems. Particularly, we discuss the benefits that can be reaped from the concept of \ac{OAM} given the quasi-optical nature of \ac{THz} systems. We also explore the role of \ac{NOMA} in \ac{THz} systems. In fact, we will see how the synergy between \ac{NOMA} and \ac{THz} bands is strengthened by the natural adoption of \ac{RIS} architectures at those frequencies.
\subsubsection{Real-time network optimization} 6G services such as \ac{XR}, holography, and digital twins necessitate an \ac{E2E} co-design of communication, control, sensing, and computing functionalities  which to date have been attempted with limited success.  Nevertheless, the inherent coupling of communication and sensing in \ac{THz} makes it plausible to hypothesize that a joint co-design of the aforementioned aspects will be  feasible and necessary. In that sense, we scrutinize the networking challenges particular to \ac{THz} systems. Subsequently, we examine novel algorithmic approaches and techniques that can be used to optimize \ac{THz} networks thus allowing them to meet the stringent requirements of beyond 5G applications. In particular, we explore the potential of \ac{AI}, particularly, the emerging concepts of generalizeable and specialized learning, as well as meta-learning in optimizing the resources of the highly varying and non-stationary \ac{THz} channel.\\
\indent After providing a panoramic exposition of the seven defining features of \ac{THz} systems, we conclude with an extensive overview of the prospective use cases. In particular, we examine the challenges and open problems of promising \ac{THz}-enabled use cases. We further underscore different ways to exploit the aforementioned defining features in each application. The four major 6G use cases for \ac{THz} systems are shown in the right hand-side of Fig.~\ref{fig:comprehensive}.\\
\indent The rest of this paper is organized as follows. The fundamentals of \ac{THz} frequency bands are discussed in Section II. Then, the seven defining features of  \ac{THz} wireless systems are developed in the subsequent sections. In Section III, the quasi-optical nature of the band is discussed. Section IV introduces our vision for \ac{THz}-tailored network architectures. Section V exposes the synergy of the \ac{THz} band with lower frequency bands. Section VI examines the possibility of performing joint sensing and communication systems at the \ac{THz} band. Section VII introduces the channel estimation and intial access process at the \ac{THz} band. Spectrum access techniques for \ac{THz} bands are proposed in Section VIII. Real-time network optimization approaches are presented in Section IX, and key \ac{THz} use cases are discussed in Section X. Finally, conclusions are drawn in Section XI.
\renewcommand{\arraystretch}{0.92}	 
\begin{table*}[t!]
	\caption{ List of our main acronyms.}
	\centering	 
	\begin{tabular}{||c | c|| c | c ||}
		\hline
		\bf{Acronym} & \bf{Description} & \bf{Acronym} & \bf{Description}\\
		\hline
		THz  & Terahertz & mmWave & Millimeter wave\\
		\hline
		SBS & Small base station & LoS & Line-of-sight\\
		\hline 
		IoE & Internet of Everything & XR & Extended reality \\ 
		\hline
		CRAS & Connected robotics and autonomous systems & NTN & Non-terrestrial network \\
		\hline
		QoPE    & Quality of physical experience & QoS & Quality of service\\
		\hline
		E2E & End-to-end & EM & Electromagnetic\\
		\hline
		GAN & Generative adversarial network & RL & Reinforcement learning \\
		\hline
		UE & User equipment& RIS & Reconfigurable intelligent surface\\
		\hline
		ML & Machine learning & VR & Virtual reality\\
		\hline
		AR & Augmented reality & MR & Mixed reality\\
		\hline
		AoI & Age of information & HRLLC & High-rate and high-reliability low latency communications \\
		\hline
		IoNT & Internet of nano-things & AoSA & Array of subarray \\
		\hline
		OFDM& Orthogonal frequency division multiplexing & OFDMA & Orthogonal frequency division multiple access\\
		\hline
		LEO & Low earth orbit & AI& Artificial intelligence\\
		\hline
		SLAM & Simultaneous localization and mapping & IS & Information shower\\
		\hline
		OAM & Orbital angular momentum & NOMA & Non-orthogonal mutliple access\\
		\hline
		UAV & Unmanned aerial vehicle& MIMO & Multiple-input and multiple-output\\
		\hline
		MAC & Medium-access-control& BS & Base station\\
		\hline
		CSI & Channel state information & TDD & Time division duplex\\
		\hline
		OMA & Orthogonal multiple access & RF & Radio frequency \\
		\hline
	\end{tabular}
\end{table*}
\section{Fundamentals of THz frequency bands}
The key advantage of communications at \ac{THz} frequencies compared to other frequency bands is the availability of an abundant bandwidth. Nevertheless, migrating towards higher frequencies tends to be naturally limited by a shorter communication range and an intermittent (on/off) link behavior. At the \ac{THz} frequencies, these  phenomena are a result of three main characteristics: a) high path and reflection losses, b) sporadic availability of \ac{LoS} links, and c) molecular absorption.  The losses in a) are naturally observed when migrating towards higher carrier frequencies. In fact, these losses are similar to the ones occuring at \ac{mmWave} frequencies, \emph{but they are more pronounced at \ac{THz} frequencies}. Nonetheless, the molecular absorption and limited narrow \ac{LoS} links have key properties that are peculiar to the \ac{THz} band. These unique \ac{THz} characteristics will be discussed next. 
\subsection{Pencil Beam LoS Links and their Repercussions}
\ac{THz} communication links are \ac{LoS} dominant. In fact, moving towards higher frequencies widens the power gap between the \ac{LoS} and the \ac{NLoS} components. Particularly, compared to the \ac{LoS} link, the power of the first-order and the second-order reflected paths are attenuated by $5-10$ $\SI{}{dB}$ and  more than $\SI{15}{dB}$ respectively as shown in \cite{lin2016terahertz} for $\SI{300}{GHz} - \SI{1}{THz}$. Furthermore, on the one hand, high attenuation losses require focusing the power of the link within a very narrow beam. On the other hand, the small footprint of \ac{THz} antennas enables transceivers to sharpen their beams and achieve beamforming gains. Thus, extremely narrow beam \ac{LoS} links, namely \emph{pencil beam \ac{LoS} links} \cite{guerboukha2020efficient}, could alleviate the attenuation losses, equip \ac{THz} with natural interference mitigation capabilities, and pave the way for overcoming the short communication range. Nonetheless, the use of narrow beams leads to new challenges related to beam tracking, beam alignment, and mobility management.\\
\indent In fact, pencil beams can be easily disrupted by blockages, a sudden deep fade, or a slight beam misalignment following any change in a user's direction. Three types of blockages arise in \ac{THz} systems: Static (buildings, trees, etc), dynamic (neighboring users), and self blockages. Static blockages are deterministic and can be reasonably modeled, in general, and neglected for indoor THz networks. However, predicting dynamic and self blockage depends on human behavior which varies based on the type of the surrounding environment. For instance, pedestrians on the road behave differently from \ac{VR} users in an indoor area. As such, a blockage model that correctly represents an intended scenario needs to faithfully capture the unique features of the considered environment, but it must not be limited to a single environmental model. Subsequently, one major challenge in modeling blockage is the \emph{generalizability} to a broad range of circumstances.\\
\indent One could naturally wonder why the use of existing solutions that work well for blockage mitigation at \ac{mmWave} frequencies \cite{zarifneshat2017protocol} is not possible at \ac{THz}. The shortcoming of such existing solutions can be attributed to three key factors, as explained next.  The communication range at \ac{mmWave} is $10\times$ larger than the \ac{THz} range, \ac{THz} \ac{LoS} beams are significantly narrower, and  \ac{THz} \ac{LoS} links are less penetrative. Hence, novel and more accurate blockage mitigation methods need to account for unique characteristics of \ac{THz} bands. For instance, blockages must be predicted within a margin of error smaller than the radii of pencil beams. Thus, there is a need for highly accurate prediction models that can characterize micro-mobility and minute changes in users orientation. Therefore, guaranteeing a sustainable \ac{LoS} \ac{THz} link is strongly intertwined with the available information on the mobility and range of motion of users. This is particularly important for future applications like holography and XR that require continuously stable THz links. Indeed, in \cite{chaccour2020can}, we showed that  that guaranteeing an \ac{LoS} link is critical for ensuring immersive \ac{XR} services at \ac{THz} bands.\\ 
\indent To alleviate the short-range nature of \ac{THz} communication links that is caused by the inevitable molecular absorption effect and high path loss, \emph{\ac{THz} could be densely deployed}. Nonetheless, given the sensitivity of beam alignment, a dense deployment could eventually create increased \ac{LoS} interference as well as significant handovers. As a matter of fact, dense networks allow users to be at a close proximity to their serving \ac{SBS} and tend to have better \emph{best-case scenarios}. Nonetheless, small cells could lead to an intermittent association between mobile users and their respective \acp{SBS}, while also jeopardizing cell-edge users performance. Thus, leading to more pronounced \emph{worst-case scenarios}.
\subsection{Molecular Absorption Effect}
At \ac{THz} frequencies, path and reflection losses are accompanied by yet another physical phenomenon detrimental to communications, the so-called \emph{molecular absorption}. This phenomenon will not only degrade the received power, but it can also intensify the noise. Thus, it introduces \emph{molecular absorption noise} in addition to the thermal noise observed at lower frequency bands. In fact, the molecular absorption effect is observed at all frequencies, nonetheless, it only has a pronounced effect at the \ac{THz} frequencies, which is why it was often neglected for lower frequencies (\ac{mmWave} and sub-6 GHz)\cite{rappaport20175g}. In fact, compared to \ac{THz}, the \ac{mmWave} frequency band exhibits richer multi-paths, has \ac{NLoS} links with a higher power, and requires wider \ac{LoS} beams. Thus, performing beam alignment and mobility management at \ac{mmWave} bands is naturally less complex than at \ac{THz} frequencies. Furthermore, molecular absorption results from the difference in energy between the higher and lower energy states experienced by the molecules of the physical medium when being transmitted. At the \ac{THz} band, molecular absorption is mainly a byproduct of water and oxygen vapor in the air. Thus, changes in meteorological conditions will lead to drastic effects on the air composition and the molecular absorption. This, in turn, makes \ac{THz} communications more suited to indoor scenarios due to a lower water vapor percentage.\\
\indent Although molecular absorption increases with the frequency, this increase is neither smooth nor monotonic. In fact, there exist some regions where a dip in the baseline of molecular absorption is observed at some carrier frequencies. Hence, some works such as \cite{rajatheva2020scoring} and \cite{rappaport2019wireless} argue that these particular windows could be targeted to benefit from their lower absorption coefficient. Nevertheless, the air composition changes based on meteorological conditions. In fact, given that the \ac{THz} molecular absorption is highly correlated to the air's water vapor, different air humidity levels will contribute to different molecular absorption levels. Hence, these windows can be potentially unreliable given their variability based on air composition. Henceforth, targeting these frequency windows makes more sense in controlled and indoor environments. Moreover, the molecular absorption baseline increases with the carrier frequency. Subsequently, at higher frequencies, more abundant bandwidth can be exploited. In turn, \emph{the molecular absorption and the large \ac{THz} bandwidth are two opposing forces.} The first is jeopardizing the performance by incurring more losses and noise, while the latter is improving the performance by boosting the capacity of the system. For example, in an indoor \ac{AR} application, we showed in \cite{chaccour2020ruin} that an increase in the bandwidth could lead to fresher \ac{AR} content and more reliable communication on average. Nonetheless, operating at higher frequencies led to a more exacerbated worst-case user scenario (in terms of content freshness) due to the molecular absorption effect.\\
\indent From a communication perspective, molecular absorption has a non-monotonic detrimental effect on the \ac{THz} links. Effectively, it adds a margin of noise, limits the communication range, and entangles outdoor opportunities. Nevertheless, the molecular absorption effect is a double-edged sword that provides several gains to the \ac{THz} sensing mechanism. In fact, the quasi-optical nature of the \ac{THz} band and its abundant bandwidth lead to additional sensing functionalities that will be discussed next. The quasi-optical nature of the \ac{THz} band is in fact the first identified THz characteristic among the seven unique defining features that we envisioned in Fig.~\ref{fig:comprehensive}.
	\begin{figure*} [t!]
	\begin{centering}
		\includegraphics[width=.75\textwidth]{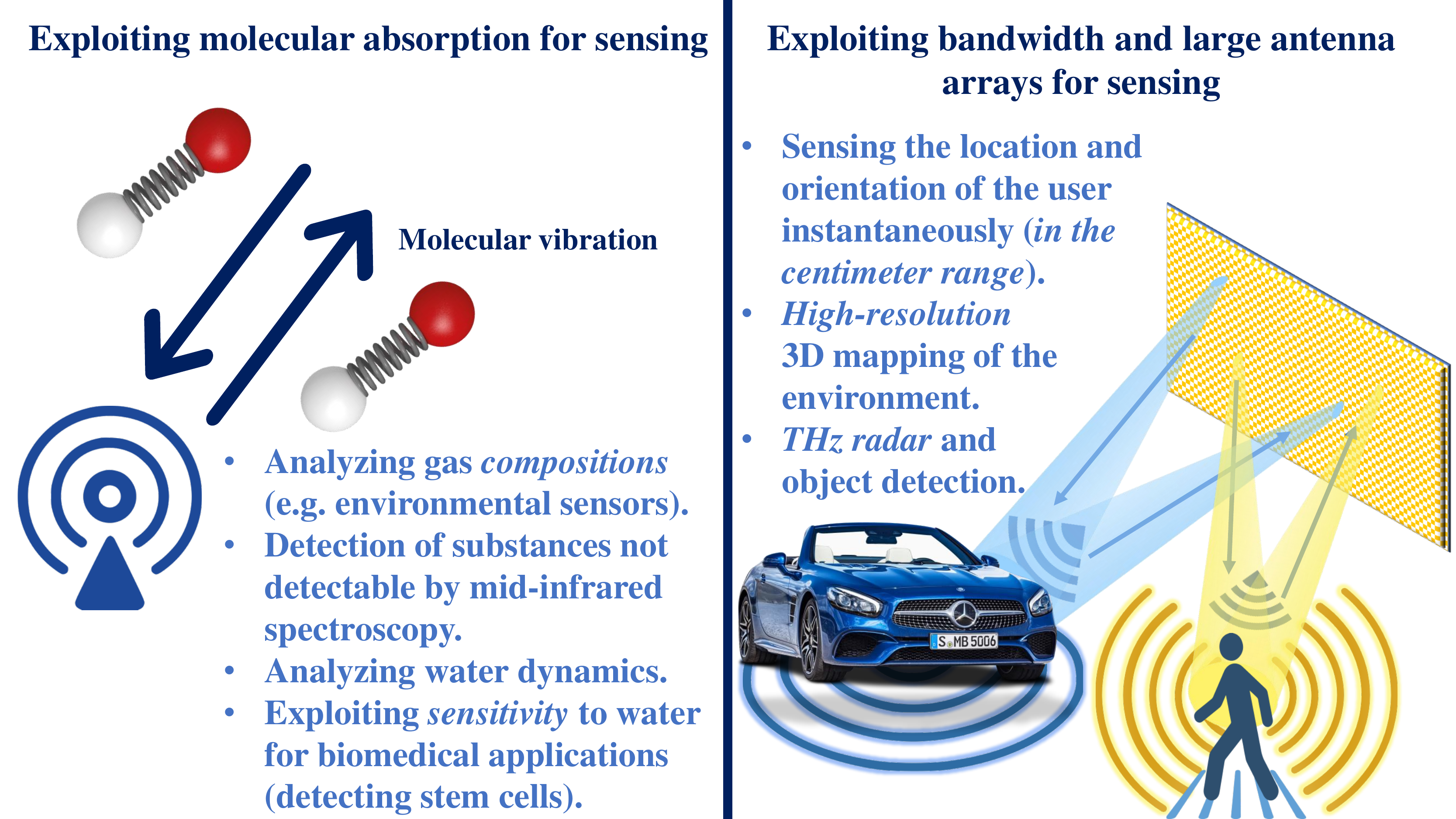}
		\caption{\small{Illustrative figure showcasing the different contexts of \ac{THz} sensing.}}
		\label{fig:sensing}
	\end{centering}
\end{figure*}
\section{Quasi-Opticality of the THz Band}
\ac{THz}'s distinct \ac{EM} nature allows it to be uniquely suited for use in \emph{wireless environmental sensing}. The first factor that unravels sensing capabilities at \ac{THz} frequencies is the molecular absorption effect as shown on the left hand-side of Fig.~\ref{fig:sensing}. In fact, although molecular absorption could hinder communication at \ac{THz} bands, this particular characteristic significantly improves \ac{THz}'s sensing capability. Effectively, molecular absorption allows \ac{THz} to have \emph{electronic smelling} capabilities, i.e. the interactions of the \ac{THz} wave with a gaseous medium result in a specific molecular absorption loss. Such a technique is known as \emph{rotational spectroscopy} \cite{sharma2016200}, and it has a high level of specificity, sensitivity, and determination of concentration. This is because this technique relies on measuring the energies of the transitions between the quantized rotational states of molecules in gases. Moreover, if more complex substances or fluid dynamics need to be identified, sensing measurements must be further processed. Therefore, after some manipulation, the measurement data can be fed to a particular neural network (based on the desired identification or classification task) to learn more entangled patterns.  The implementation of such learning mechanisms for sensing is an open research area. This sensing potential has seen light in environmental nano-sensors for pollution monitoring. \ac{THz} frequency bands are capable of sensing the chemical compounds present in the environment with one part per billion density as shown in \cite{tekbiyik2019terahertz}. Hence, examining the interactions of the molecular absorption paves the way for many revolutionary sensing applications that require an extremely high accuracy and precision. As a matter of fact, beyond environmental sensors, this functionality can be beneficial to the medical field in procedures such as cancer diagnosis. For instance, \ac{THz} radiation has the potential to non-invasively detect stem cells \cite{yu2019medical} and lead to a major leap in the diagnostic. \\
\indent Meanwhile, for high-resolution environmental sensing\footnote{It is important to note that the terms imaging and sensing have been used interchangeably by the literature to denote personal radars and the ability to illuminate objects and scenes with \ac{THz} waves to capture their reflection and angular precision. For consistency, given that \emph{sensing measurements} are being procured, throughout this work, this functionality will be referred to as sensing.} at a larger scale, the usage of large arrays coupled with \ac{THz}'s abundant bandwidth brings forth additional key sensing capabilities. These capabilities include high-resolution \ac{THz} radar, accurate \ac{UE} localization \cite{sarieddeen2020next}, precise 3D mapping (e.g. acting as a personal radar in indoor areas), and minute device/user orientation sensing. \ac{THz}'s different sensing contexts and capabilities are illustrated in the right hand-side of Fig.~\ref{fig:sensing}. In all of these processes, sensing parameters from the transmitted or the reflected \ac{THz} \ac{EM} wave are measured and examined. These parameters include but are not limited to: time delay, angle-of-arrival, angle-of-departure, Doppler frequency, and physical patterns of objects detected \cite{rahman2020enabling}. Subsequently, the sensing mechanism at \ac{THz} frequencies can provide active \acp{BS} with \emph{situational awareness}, i.e., instantaneous location and orientation information at the centimeter level and in 3D space of nearby \acp{UE}. This situational awareness provides a new breed of data that can be exploited in data-driven \ac{THz} networks for effective beam management, mobility, and blockage avoidance. Similar to the \ac{THz}'s band communication functionality that exhibits a high-rate but is limited by a short range, \ac{THz}'s sensing capability has a high resolution but its detection ranges are short. Thus, sensing at \ac{THz} frequencies is apropos for capturing subtle user movements by characterizing the range of motion of \acp{UE} within errors in the centimeter range, i.e., micro-mobility and micro-orientation changes. This is clearly useful in the context of many 6G services, including \ac{XR} and holography (discussed in Section X-A) that necessitate a continuous \ac{SLAM} input to guarantee an immersive experience. \\
\indent Thus, on the one hand, wireless \ac{THz} systems can provide communication services with unprecedented high rates that are challenged by the \ac{THz} band's unique channel impediments. On the other hand, \ac{THz} systems can be used as a high-resolution sensing facility limited by a short range. Effectively, the quasi-optical nature of the \ac{THz} band brings to light a fellowship between \emph{high-rate communication and high-resolution sensing} capabilities. These capabilities are the core assets of \ac{THz} systems, and they can eventually make them superior to the lower \ac{RF} bands as well as the higher optical bands. In fact, \emph{each blockage of a communication link is a sensing opportunity}. Hence, to ensure a strategic deployment of such integrated communication and sensing functionalities, we next discuss effective architectures for wireless \ac{THz} systems.  
\section{How to Strategically Deploy THz Networks?}
Migrating towards higher frequencies such as \ac{THz} overcomes the bandwidth scarcity challenge, but it leads to highly varying and uncertain channels. For instance, the distinct physical characteristics of \ac{THz} outlined in Section II and III mandate a novel network architecture capable of sustaining the traffic capacity and connection density requirements, as well as the multi-function nature of \ac{THz} systems. In particular, it is important to design the underlying architecture in a way to take advantage of the spectrum supremacy of \ac{THz} frequency bands and enable joint communication and sensing while mitigating their uncertain channel and intermittent nature.
	\begin{figure*} [t!]
	\begin{centering}
		\includegraphics[width=.8\textwidth]{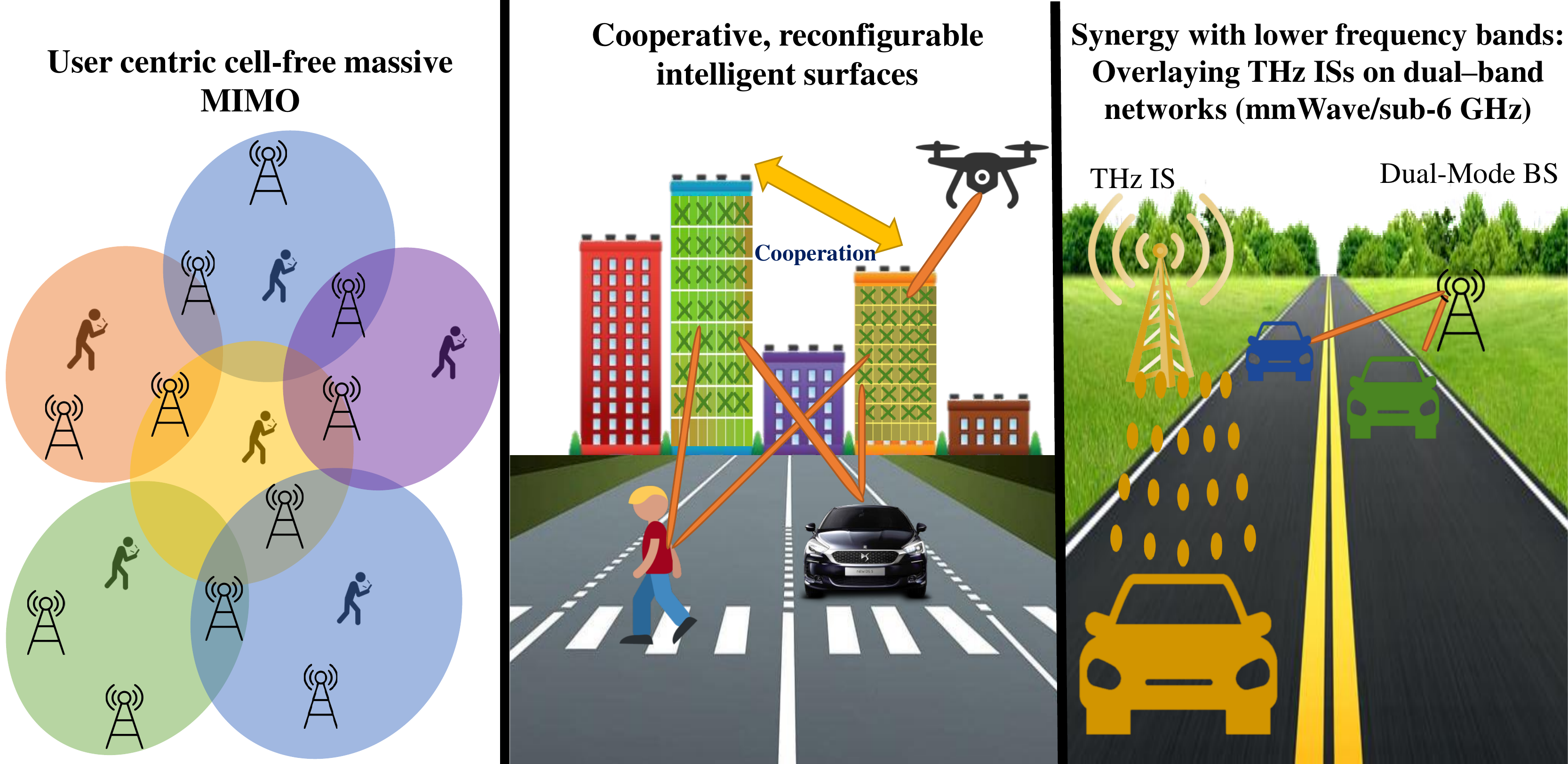}
		\caption{\small{
			Illustrative figure showcasing \ac{THz} enabling architectures and the synergy with lower frequency bands. The deployment of \ac{THz} of wireless networks should bring out novel cell less architectures, rely on cooperative \acp{RIS}, and synergistically coexist with lower frequency bands.}}
		\label{fig:architecture_thz}
	\end{centering}
\end{figure*}
\subsection{From Ultra Massive MIMO to Cell-Free Massive MIMO}
\subsubsection{Motivation}
\indent Leveraging the \ac{THz} small antenna footprint, allows us to embed ultra massive \ac{MIMO} systems within a few square millimeters thus providing pencil beamforming in a scalable and efficient way \cite{sarieddeen2020next}. Furthermore, adopting an \ac{AoSA} architecture allows dividing large antenna arrays into multiple sub-arrays, which helps in improving not only the beamforming gain but also the energy efficiency \cite{faisal2019ultra}. Indeed, massive \ac{MIMO} was a staple of 5G systems, and, similarly, in beyond 5G systems, ultra-massive \ac{MIMO} can potentially provide \ac{THz} communications with improved beamforming, energy efficiency, and multiplexing benefits. Nonetheless, the true success of ultra massive \ac{MIMO} for sub-6 GHz was a result of channel hardening and favorable propagation. These phenomena occur with Rayleigh fading channels having \ac{i.i.d.} channel coefficients. However, \ac{THz} frequency bands are dominated by \ac{LoS} links and characterized by very sparse and low rank channels. In fact, the number of \ac{NLoS} links decreases as we increase the carrier frequency of operation. Also, while a few \ac{NLoS} links exist in the channel matrix, capitalizing them becomes an ineffective task due their weak received power at the \ac{UE}. Hence, as a result of the peculiarity of the \ac{THz} channel, our ability to leverage these attractive phenomena of massive \ac{MIMO} is questionable. For instance, when deploying massive \ac{MIMO} at lower frequency bands, the \emph{energy efficiency-capacity} tradeoff is an important metric in the system design. Primarily, this tradeoff is determined by the \ac{MIMO} technique used: On the one hand, spatial multiplexing is capable of increasing spectral efficiency linearly with the number of transmit antennas, however, that comes at the expense of a higher energy consumption and complexity. On the other hand, spatial modulation is limited by increasing the spectral efficiency by base two logarithm of the number of transmit antennas. However, spatial modulation enables a system with a higher energy efficiency, lower complexity, and cost, given that one transmit antenna is active at a time. Meanwhile, at \ac{THz} frequency bands, the dynamics of this tradeoff are slightly different. First, \ac{THz} frequency bands naturally provide a high data rate, and thus the need to expand the capacity and spectral efficiency further is less motivated than at lower frequency bands. Second, given that \ac{THz} bands have a small antenna footprint, the number of antennas can be made very large at low costs \cite{sarieddeen2019terahertz}. As such, spatial modulation's system efficiency is higher at \ac{THz} frequency bands. Nonetheless, it is worth mentioning that hardware impairments need to be considered in the analysis of spatial modulation at \ac{THz} frequency bands as done in \cite{mao2020spatial}. Indeed, the works in \cite{sarieddeen2019terahertz} and \cite{mao2020spatial} should be followed by research that investigates ultra-massive spatial modulation \ac{MIMO} at \ac{THz} frequency bands under non-idealized channel conditions and assumptions.\\
\indent Furthermore, at \ac{THz} frequencies, network densification with massive \ac{MIMO} small cells is a necessary technique to ensure seamless coverage despite the short communication range. Subsequently, this phenomenon raises multiple challenges such as, intermittent connectivity of cell edge users, increased inter-cell interference, and significant overhead with global \ac{CSI}. Here, we note that, although massive \ac{MIMO} solutions have been proposed for \ac{mmWave} networks \cite{mumtaz2016mmwave} and \cite{rappaport2017overview}, these solutions are not directly applicable to the \ac{THz} frequency bands due to the poorer multi-path propagation and the sparser, lower-rank \ac{THz} channels.
\subsubsection{Opportunities}
\indent Given the challenges of using dense \ac{THz} massive \ac{MIMO} networks and the fact that \ac{THz} cannot reap all the benefits of ultra massive \ac{MIMO}, we naturally pose the following question: \emph{Can we engineer an opportunistic, \ac{THz} tailored architecture that reaps more benefits from THz's disposition?}\\
\indent An evident answer to this question would be to untie the architecture from cellular boundaries. In that sense, we can think of leveraging the concept of cell-free massive \ac{MIMO} \cite{ngo2017cell} that is viewed as an effective approach to address the challenges brought by dense massive \ac{MIMO} small cells. Cell-free massive \ac{MIMO} could, in principle, provide a uniform experience across all users and improve personalized \ac{QoE} without jeopardizing the experience of cell-edge users. As such, it allows us to close the gap between best-case and worst-case performance. This is particularly beneficial for \ac{THz} systems given the considerable gap between the best-case and worst-case user performance.\\
\indent We now recall that cell-free massive \ac{MIMO} only requires local \ac{CSI} at each \ac{BS} in contrast to the global \ac{CSI} needed in massive \ac{MIMO}. This, in turn, could significantly reduce the overhead and improve the reliability and latency at the \ac{THz} band. As such, this allows mitigating one of the challenges pertaining to the \ac{THz} channel estimation process. However, as explained in Section VII, this process suffers from multiple other challenges that are independent of the architecture adopted.\\
\indent Furthermore, in cell-free massive \ac{MIMO}, one can exploit the concept of user-centric clustering to constrain the number of \acp{UE} that can be served per \ac{BS} without creating cell boundaries. Such a clustering approach, consists in deploying dynamic and possibly overlapping clusters of \acp{BS} based on the needs of the user, as shown in Fig.~\ref{fig:architecture_thz}. In dense \ac{THz} deployments, clustering can potentially suppress inter-cell interference and connect multiple cooperative \acp{BS} to a given user. This is uniquely beneficial for \ac{THz} systems given their short range of communication, their intermittent links, and their high likelihood of handover failures. As such, having multiple active links enhances the \ac{THz} link reliability and continuity. Clearly, cell-free massive \ac{MIMO} can potentially improve the availability of \ac{LoS} \ac{THz} links, mitigate interference, as well as reduce frequent handover, handover failures, and \ac{CSI} overhead. 
\subsubsection{Challenges}
\indent \\$\bullet$ 
\emph{Scalability:} Despite the aforementioned benefits of cell-free massive \ac{MIMO} for \ac{THz} systems, the scalability challenge of cell-free architectures remains a partially solved problem. In \cite{interdonato2019scalability}, the authors addressed some system scalability challenges of cell-free massive \ac{MIMO} by proposing a scheme in which a user is served by all cell-centric clusters related to a given user-centric cluster. However, such solutions are still at their nascent stage and more realistic channel models that include channel correlations and consider multi-antenna \ac{UE} are needed. Furthermore, user-centric cell-free massive \ac{MIMO} requires a cooperative, dynamic, and time variant clustering mechanism, which is another key challenge. When the size of the network grows, one could investigate the open problem of providing a scalable network cooperation to these clusters.\\
$\bullet$ \emph{Coherent Processing:} The joint coherent processing among \acp{BS} will be highly demanding in terms of rate and computation; yet, the demands for high-capacity and high data rates can be met given the operation at \ac{THz} frequencies. Nevertheless, this process could still lead to a significant amount of processing and signaling overhead (e.g. power control, synchronization, pilot assignment) that will be needed to reliably exchange \ac{CSI}. Overcoming such challenges requires novel channel estimation techniques that are designed to cope with both the cell-free architecture as well as the highly varying \ac{THz} channel simultaneously. These methods will be explained in Section VII.
\subsection{Cooperative Reconfigurable Intelligent Surfaces}
\subsubsection{Motivation}
Since the \ac{THz} performance is contingent on the availability of \ac{LoS} connections \cite{chaccour2020risk} and \cite{chaccour2020can}, it would be beneficial if one can to customize the wireless propagation environment in a way to guarantee a continuous \ac{LoS} link. Here, the concept of \acp{RIS} can be exploited to carefully re-engineer the propagation environment.  \acp{RIS} are surfaces of electromagnetic material that can be electronically controlled by integrated circuits \cite{basar2019wireless}. In fact, \acp{RIS} can potentially transform wireless environments into software-defined platforms thus providing enhanced beam and mobility management over the uncertain \ac{THz} channel. To increase the level of control exerted, \emph{multiple \acp{RIS} could be used to cooperatively sense the environment and connect with narrow \ac{THz} beams to provide continuous and non-intermittent communication links to users} as shown in Fig.~\ref{fig:architecture_thz}.
\subsubsection{Opportunities in using Holographic \acp{RIS} for THz frequency bands} Owing to the small footprint of \ac{HF} band transmissions and \ac{THz} communications, a new breed of intelligent surfaces, dubbed \emph{holographic} \acp{RIS} \cite{wan2020terahertz} has come to light. This new concept, integrates a very large number of small antenna elements to realize a holographic array having a spatially \emph{continuous} aperture. Such structures can be viewed as a theoretically infinite number of antennas, i.e., an asymptotic limit of ultra massive \ac{MIMO}. As such, they can potentially outperform non-holographic \acp{RIS} given the increase in the number of metasurfaces, by achieving a higher spatial resolution, and
enabling the transmission and detection of \ac{EM} waves with arbitrary spatial frequency components \cite{huang2020holographic}. Particular to \ac{THz} networks, \emph{the detection capability of a holographic \ac{RIS} can provide a better sensing system}\footnote{Enhancing sensing using \acp{RIS} entails improving the localization, radar, and situational awareness capability of \ac{THz}.}, which is an essential process for the assessment of a wireless user's orientation and position \emph{instantaneously}. Additionally, given the small number of propagation paths between a \ac{THz} \ac{SBS} and a \ac{UE}, having multiple \acp{RIS} mimics the behavior of a multi-path propagation environment and converts limited \ac{LoS} channel models into richer ones, thus improving spatial multiplexing capabilities. \acp{RIS} are also particularly suitable for sensing because they can optimize and program their metasurface configurations. This allows generating a massive number of independent paths to enhance the object detection and localization process. Indeed, holographic RISs can enable precise 3D environmental mapping through the inherent wireless sensing capabilities of the THz bands that we discussed in Section III and VI.
\subsubsection{Opportunities in Near-field \ac{RIS} communication at \ac{THz}}
\indent Given the dense \ac{THz} architecture, the high number of metasurfaces, and the operation at a short range, \acp{RIS} can potentially satisfy the \emph{co-phase condition} and operate in the near-field \cite{liu2020reconfigurable}. Near-field communications allow focusing the reflected beams towards a new focal point, in contrast to far-field communications that only provide anomalous reflections to the incident wave. Furthermore, near-field communication allows the network to exploit wavefront curvature to reduce the need for infrastructure and synchronization. Thus, due to the shorter range of communication and network conditions at \ac{THz}, \acp{RIS} can first use their beamforming capability by boosting links suffering from low power. Second, they create a virtual \ac{LoS} path to overcome blockages that constitute a big challenge in the \ac{THz} operation. Third, the phase of arrival measurement of the transmitted signal obtained from near-field propagation opens the door for better sensing capabilities using \acp{RIS}, as it allows us to obtain information about the angle and the distance to an \ac{RIS}. This information obtained along with the time of arrival of the signal can be further exploited to determine unknown clock biases and spherical wave localization \cite{wymeersch2020radio}. Henceforth, novel channel models and signal processing methods are needed in order to conform with the \ac{THz}-\ac{RIS} near-field propagation environment and to characterize their behavior accurately.
\subsubsection{Challenges}
\indent   \\$\bullet$ \emph{Complex Cooperation: }Simultaneously employing multiple holographic \acp{RIS} imposes novel challenges on the network due to the complexity of enabling their cooperation. For instance, having a cooperative \ac{RIS} system makes it infeasible to use channel acquisition schemes that rely on penalizing the achievable rates to account for pilot overhead from channel estimation \cite{wymeersch2020radio}. Furthermore, an \ac{RIS} acting as an intelligent reflective surface cannot process and estimate the channel without dedicated receive chains. Thus, such a compound channel will exacerbate the overhead stemming from the channel estimation and processing delays of the network. As some of the beyond-5G services have \emph{very strict latency requirements}, i.e., a near-zero \ac{E2E} latency, cooperation and scheduling among multiple \acp{RIS} must operate in a stringent real-time fashion. Nonetheless, relying on partial \ac{THz} \ac{CSI}, i.e., statistical information of the channel such as spatial, frequency, and path correlation or properties of the channel matrix, hinders the high reliability of links. \\
\indent $\bullet$ \emph{Accurate \ac{CSI} acquisition:} Obtaining accurate \ac{CSI} for multiple \acp{RIS} is a key challenge. First, it involves the estimation of multiple channels simultaneously. Second, \ac{THz} \acp{RIS} will have a very large number of metasurfaces\footnote{This phenomenon is more pronounced with holographic \acp{RIS}}, each of which might have particular non-linear hardware characteristics. Third, the highly-varying \ac{THz} channel limits the \ac{CSI}'s validity to very short period of time. One potential solution to the aforementioned challenges is to exploit the sparse nature of the channel and extract features from the geometric configuration of the network to learn more accurate \ac{CSI}.\\ 
\indent $\bullet$ \emph{Data-driven methods:} Exploiting the aforementioned channel features calls for data-driven and proactive predictive methods, capable of bypassing the highly-varying \ac{THz} channel and inferring accurate \ac{CSI} within the coherence time of \ac{THz} and the stringent \ac{E2E} latency requirements of 6G applications. These methods will be further developed in Section VII. We note that, for all considered architectures one can also envision several enhancements to further overcome the unique \ac{THz} challenges. For example, in order to further improve the coverage range, one can consider the use of multi-hop directional transmissions as discussed in \cite{ahmadi2021reinforcement}.\\
\indent After comprehensively discussing the potential \ac{THz} architectures, their specific challenges, the opportunities that they present, and the open research problems that need to be examined; we next discuss the synergy of the \ac{THz} band with lower frequency bands and its role in the deployment of future wireless systems.
\renewcommand{\arraystretch}{1.30}
\begin{table*}[t]
	\footnotesize
	\caption{ Wireless cellular frequency bands rivalry and complementarity.}
	\centering
	\begin{tabular}{|p{2.5cm} | p{4cm} | p{4.5cm} | p{4.5cm}|}
		\hline
		& \textbf{Sub-6 GHz} & \textbf{mmWave} ($30- \SI{100}{GHz}$)& \textbf{ sub-THz and THz} ($0.1- \SI{10}{THz}$) \\
		\hline
		\textbf{Distance}  & High range & Medium to short range ($\leq \SI{200}{m}$) & Short range  ($\leq \SI{20}{m}$)\\
		\hline
		\textbf{Bandwidth}  & Limited & Medium to Large 	&Large\\ 
		\hline
		\textbf{Data rates} & Limited &  Medium to High (up to $\SI{10}{Gbps}$) & High  (up to $\SI{100}{Gbps})$\\
		\hline
		\textbf{Interference}  & Mitigated by techniques like OFDM and OFDMA & Mitigated by beamforming
		& Mitigated by sharp pencil beamforming	\\
		\hline
		\textbf{Noise source} & Thermal noise & Thermal noise & Molecular absorption noise and Thermal noise\\
		\hline
		\textbf{Blockage} & Not susceptible & Susceptible & Highly susceptible \\
		\hline
		\textbf{Beamforming} & Medium to narrow beams & Narrow beams & Very narrow beams \\
		\hline
		\textbf{Horizons to explore} & Expanding the midband coverage & NLoS communications and long-range sensing functions & Reliable and low latency communications, integrated sensing and communication systems\\
		\hline
		\textbf{Viable architectures} & Massive MIMO & Ultra massive MIMO, RIS, and UAV-RIS & Cell-free massive MIMO and holographic intelligent surfaces\\
		\hline
		\textbf{Significant caveats} & Low rates and spectrum inefficient & Susceptibility to mobility and blockages & Susceptibility to micro-mobility, orientation, air composition and blockages  \\
		\hline
		\textbf{Applications} & Low-rate and latency tolerant services& Vehicular networks, radar, UAVs, and IoT & XR, holography, IoE, NTNs, sensing, and nanosensors\\
		\hline
	\end{tabular}
	\label{table:Rivalry}
\end{table*}
\section{Synergy with Lower Frequency Bands}
\subsection{Motivation}
Independent of the deployed architecture, integrating \ac{THz} communications with \ac{mmWave} and sub-6 GHz bands provides many opportunities. Effectively, investing on this synergy between the \ac{THz} band and lower freqeuncy bands allows future wireless systems to achieve a realistic universal coverage, and to deliver more scalable network solutions. Particularly, such an integration allows us to exploit the benefits of \ac{THz} communications and sensing capabilities in outdoor scenarios as well as to service highly mobile \ac{UE}, despite the short communication distance of \ac{THz} (due to severe power limitations and propagation losses). For example, highly mobile \ac{CRAS} applications like autonomous vehicles or even drones require downloading high-quality 3D maps instantaneously. Providing enhanced data rates to this process can be done by exploring the use of \ac{THz} \ac{SBS} as enablers of an \emph{\ac{IS}} as illustrated in Fig.~\ref{fig:architecture_thz}. Conversely, blockage-prone short-range \ac{THz} links can benefit from the exchange of control information at the lower frequency bands thus facilitating a reliable real-time reconfiguration of \ac{THz} networks\\
\indent Similarly, this integration can be done by deploying dual-band \ac{THz} and \ac{mmWave} \acp{SBS}, or even triple-band \acp{SBS}. Meanwhile, we expect \ac{THz} \acp{IS} to be the first stage in the evolution of this foreseen frequency band coexistence (since this will only require overlaying new \ac{THz} operated \acp{SBS} to existing dual-band networks). This \ac{IS} deployment provides accessibility to \ac{THz} in current \ac{mmWave} and sub-6 GHz networks and allows serving network slices that require extremely high rates that can only be satisfied by \ac{THz}. In the next stage, we envision more networks to rely on dual-band \ac{mmWave} and \ac{THz} frequencies, in a way to provide high rates and extremely high rates, respectively. Such a deployment will see light after the \ac{IS} stage as a result of: a) The major infrastructure changes needed to achieve a ubiquitous coverage of dual-band \ac{mmWave}-\ac{THz} \acp{SBS}, b) The maturity of a higher number of applications that require extremely high data rates, and c) The compliance of \ac{mmWave}-\ac{THz} \acp{SBS} with existing wireless standards and network elements.
\subsection{Opportunities}
\indent Integrating the \ac{THz} frequencies with lower frequency bands (\ac{mmWave} and sub-$\SI{6}{GHz}$) allows next-generation wireless systems to seize the following opportunities:
\begin{itemize}
\item Improve the reliability and continuity of \ac{THz} communication links by enhancing blockage prediction and exchanging control information using the links with larger and reliable range over the lower frequency bands.
\item Incorporate sensing mechanisms that are capable of providing high-resolution information at short range using \ac{THz} bands and radar-like information at long ranges using \ac{mmWave} bands. The concept of joint sensing and communication will be elaborated in Section VI.
\item Benefit from the spatial correlation between the three frequency bands and develop knowledge for traffic scheduling and training overhead reduction \cite{alrabeiah2019deep}.
\item Provide a versatile range of network slices for 6G systems with different rate, latency, and synchronization by selecting unique communication band support and smoothly handing off users from one band to another.
\item Prefetch high-rate demanding data such as \ac{AR} content using \acp{IS} while supporting conventional lower rate services using the existing \acp{SBS}.
\end{itemize}
Clearly, integrating \ac{THz} frequency bands with \ac{mmWave} and sub-$\SI{6}{GHz}$ facilitates its introduction to current outdoor wireless networks, improves its link reliability, and extends its coverage. However, differences in the \ac{EM} behavior, signal propagation properties, and available bandwidths lead to significant differences in the achievable rate, reliability, and latency. Such differences and heterogeneity lead to many challenges and open problems as explained next.
\subsection{Challenges and Open Problems}
\indent Integrating the \ac{THz} band with lower frequency bands (\ac{mmWave} and sub-6 GHz) \acp{BS} \cite{semiari2017joint, semiari2017caching, coll2019sub}  hybridizes the network and enhances the reliability of its links. While such hybridization improves the coverage, reliability, and provides more versatile services, it has also lead to multiple novel challenges. \\
	\indent $\bullet$ \emph{Hybridization Techniques:} When using heterogeneous frequency bands, a key challenge is to implement new schemes that allocate interfaces between the three (or more) different frequency bands. One way to implement this hybridization efficiently, is to deploy \emph{time-critical and vitally-robust} interfaces at lower frequency bands to support beamforming, initial access, and channel estimation. This, in turn, reduces the training overhead associated with beamforming and channel estimation at \ac{HF} bands particularly \ac{THz}. Similarly, heavy data transmission interfaces must rely on bandwidth-abundant \ac{THz} \acp{IS} to enhance the data rates. Nonetheless, orchestrating control, time-critical, and data-hungry information over multiple frequency bands while maintaining a high \ac{QoS} and \ac{QoE} necessitates novel network management techniques. In fact, to propose efficient resource and network management schemes, that reinforce the synergy between \ac{THz} and lower frequency bands, one can leverage the data collected from \acp{BS} and \acp{RIS}. As a result, this allows us to learn the complexities entailed by the alliance of versatile frequency bands used. Indeed, achieving a successful hybridization is an open research area that opens the door for novel \ac{ML} and \ac{AI} techniques capable of learning the most effective interface allocation, while optimizing the cooperation between the heterogeneous frequency bands.\\
	\indent $\bullet$ \emph{Effective Control:} Another key challenge of integrating multiple frequency bands is to devise effective control and signaling protocols that can handle the different characteristics of the different bands (e.g., see \cite{semiari2019integrated} and \cite{shokri2015millimeter} for the case of \ac{mmWave} $-$ sub-$\SI{6}{GHz}$ band integration).\\
	\indent $\bullet$ \emph{THz \acp{IS}:} There are several challenges that are peculiar to the use of THz \acp{IS}. For instance, the fixed location of \acp{IS} might limit the ability of \acp{UE} to prefetch high-rate content or experience very high data rates. To address this challenge, deploying denser dual-band \ac{mmWave}-\ac{THz} \acp{SBS} increases the likelihood of the delivery of a high-rate \ac{THz} connection to a user. Nevertheless, this deployment leads to higher infrastructure costs and an increased complexity.\\
	\indent $\bullet$ \emph{Open Problems:} The coexistence of multiple frequency bands brings forward several additional open problems such as: a) Mapping control information and payload to different frequency bands, while taking into account their causal relationship, i.e. the fact that data can only be interpreted if the appropriate control information is in place, b) Joint scheduling and spectrum management to make use of multi-connectivity across bands and maintain high reliability, c) Associating users to cells while maintaining a load balance, and d) Learning the \ac{THz} network conditions by leveraging data and information from lower frequency bands. \\
	\indent It is worth noting that handling and solving such problems efficiently requires capitalizing on the distinct features and complementarities of these frequency bands as shown in Table~\ref{table:Rivalry}. Clearly, a successful deployment of wireless \ac{THz} networks relies on adopting the aforementioned architectures and investing in the coexistence with lower frequency bands. Effectively, to take advantage of all the benefits of the quasi-optical behavior, such networks need to migrate towards \emph{versatile wireless systems} that can perform multiple functions such as sensing, communications, and localization, among others. Thus, we next discuss the prominent role of joint sensing and communication systems.
\section{Joint Sensing and Communication Systems}
Wireless systems are rapidly evolving from solely relying on communication services to versatile systems with joint sensing and communication capabilities \cite{rahman2020enabling}. This transformation is a byproduct of two main factors. \emph{First}, future wireless services necessitate some form of sensing input as part of the functionality of the application. For example, autonomous vehicles necessitate radar capabilities that enable them to avoid collisions. \emph{Second}, equipping systems with a joint sensing and communication capability leads to more effective:  \begin{enumerate}[label=(\roman*)]
	\item Sensing served by communication feedback. \item Communication served by sensing feedback. \item Mutual sensing and communication feedback to fulfill joint goals.\end{enumerate} These abilities bring  an \emph{interesting  fellowship of communication and sensing, as each communication blockage represents  a  sensing  opportunity  and  vice  versa}. In fact, \ac{THz}'s sensing outcomes are versatile (e.g. object detection, \ac{UE} tracking, 3D mapping) as previously shown in Fig.~\ref{fig:sensing}. Meanwhile, the common denominator of these outcomes is the ability to analyze the transmitted or reflected \ac{THz} \ac{EM} wave in a way to infer measurements. Such measurements include the angle-of-arrival, angle-of-departure, time-of-arrival, Doppler frequency, and physical patterns of objects detected. Collecting and processing these high-resolution measurements enable \ac{THz} systems to localize \ac{UE} and detect objects in the $\SI{}{cm}$ and $^{\circ}$ range for example. 
\subsection{Opportunities and Advantages}
\indent Incorporating integrated sensing and communication functions in future wireless \ac{THz} systems holds under its belt many opportunities and advantages:
\begin{itemize}
	\item \emph{Improved pencil beamforming:} Maintaining reliable pencil \ac{THz} beams has always been a difficult process for highly mobile environments. Having readily available and continuous sensing feedback to build a situational awareness leads to a better initial access, an improved beam tracking, and enhanced user association.
	\item \emph{Predictive resource usage:} Sensing paves the way for situational awareness and the means to characterize environmental changes and user gestures. Such actions are usually followed by a certain communication demand in services like \ac{XR}. Subsequently, the sensing input can be used to make predictive allocation of communication resources.
	\item \emph{Spectrum efficiency and coexistence:} Allowing \ac{THz} sensing and communications functionalities to dynamically share the spectrum inherently increases the spectrum efficiency. For instance, if sensing and communication functionalities do not need to operate simultaneously, a significantly improved spectral efficiency would be observed compared to disjoint systems. In essence, spectrum sharing is an option for the coexistence of sensing and communication functionalities given \ac{THz}'s ample bandwidth. Alternatively, one can benefit from \ac{THz}'s quasi-optical nature and exploit techniques like \ac{OAM} to multiplex sensing and communication functionalities. \ac{OAM} will further be elaborated in Section VIII.
	\item \emph{New use cases:} Benefiting from the sensing feedback makes \ac{THz} systems an ideal candidate to many applications requiring a dual sensing and communication feedback. For example, \ac{XR} services require a network that can provide an instantaneous high-resolution localization of the user as well as a high data rates. Herein, \ac{THz} systems can jointly serve such a use-case especially in indoor areas. 
	\item \emph{Cost efficiency:} Integrating the communication and sensing functionality on a single platform leads to reduced costs and size \cite{rahman2020enabling}.	
\end{itemize}
Henceforth, joint sensing and communications systems provide ample opportunities for innovative wireless systems. Nonetheless, the short communication range and the intermittent links, on the one hand limit the reliability and coverage of communication links. On the other hand, they limit the sensing scope and the situational awareness coverage. Next, we will discuss the means to improve joint sensing and communication systems.
\subsection{Effective Strategies for Joint Sensing and Communications}
The overall performance of joint sensing and communication systems is primarily contingent upon improving the link reliability, continuity, and coverage. Incorporating systems that have multiple independent paths extends the communication reliability and improves the sensing richness, i.e., measurements become capable of capturing longer range objects and larger number of details in terms of situational awareness. Attaining these goals can be performed by capitalizing on the role of \acp{RIS} and on the synergies between the lower frequency bands and \ac{THz}. The roles and challenges accompanying such effective strategies include:
\begin{itemize}
\item \emph{Role of lower frequency bands:} Extending the sensing capability for longer ranges can be achieved by integrating \ac{THz} and \ac{mmWave}. This combination allows establishing sensing services with more versatile functionalities. For instance, this allows us to extend the joint sensing and communication capability to outdoor applications for services that require radar and environment sensing like autonomous vehicles. Nonetheless, this gives rise to a tradeoff between the resolution achieved and the range of sensing. Hence, having a continuous feedback between communication and sensing signals diminishes the uncertainty surrounding \ac{HF} bands like \ac{THz} and \ac{mmWave}. For instance, here, the \ac{THz} sensing input potentially detects closer objects with higher precision. Moreover, \ac{mmWave}'s sensing input can better sense farther objects with a lower resolution. Combining this dual sensing input could provide a better situational awareness and an improved blockage mitigation for both \ac{THz} and \ac{mmWave}. While such a coexistence improves the joint system's capability, it leads to a complexity in the deployment. The challenges surrounding the coexistence with lower frequency bands are explained in Section V-C.
\item \emph{Role of \acp{RIS}:} The use of \ac{RIS}-enhanced architectures could further improve the \ac{THz} sensing performance by enabling the creation of multiple independent paths carrying out richer information about the dynamics of the environment \cite{hu2019reconfigurable}. Particularly, \acp{RIS} allow us to perform the sensing process by reflecting the intended signals in precise directions, i.e., by adjusting their metasurfaces via a controller without consuming any additional radio resources. Nonetheless, \acp{RIS} acting as an intelligent reflector are limited by their inability to emit sensing or pilot signals to initiate an active sensing process.\footnote{Active sensing refers to the process of emitting radiation in the direction of a desired target to be investigated. An active sensor detects and measures the radiation that is reflected or backscattered from the target. Passive sensing, on the other hand, detects radiation that is emitted or reflected by the object or scene being observed. In this case, \acp{RIS} will be detecting the \ac{THz} radiation emitted by \acp{SBS}.} In contrast, \acp{RIS} acting as transceivers \cite{jung2020performance, jung2019reliability, jung2019performance} can send pilot signals, have an improved processing capability, and could perform both active and passive sensing. Nonetheless, their deployment comes with an increased infrastructure and energy cost. Thus, selecting the mode of operation of \acp{RIS}, i.e., as intelligent reflectors vs. transceivers, will lead to an \emph{energy efficiency $-$ processing capability} tradeoff. Here, intelligent passive reflectors can provide sensing and communication control to the network operator with a high energy efficiency, but at a low processing and transmission capability. Meanwhile, transceivers can better transmit and process sensing and communication \ac{EM} waves, at the cost of increased complexity and a reduced energy efficiency. 
\end{itemize}
After discussing the deployment strategies and integrated frequency bands techniques that can reap the benefits of joint sensing and communication systems, we next discuss the challenges in employing joint sensing and communication systems at \ac{THz} bands.
\subsection{Challenges}
While joint sensing and communications systems present many opportunities, they also give rise to several new challenges that must be addressed:\begin{itemize} 
	\item \emph{Resource sharing and allocation:} The resources (e.g. time, space, frequency) that must be used for sensing and communication signals can be statically or dynamically shared. Dynamic sharing of spatial resources (e.g. number of antennas or metasurfaces allocated) is more efficient, however, \ac{THz} communication beams must be stable and pointing sharply, whereas \ac{THz} sensing requires time-varying directional scanning beams. The design of scheduling and dynamic resource allocation schemes while balancing the \emph{high data rate and high-resolution sensing tradeoff} is therefore an open research area.
	\item \emph{Coexistence schemes:} Sensing and communication can co-exist throughout different approaches such as coexistence in spectral overlap, coexistence via cognition, or through functional coexistence  \cite{zheng2019radar}. Each of these coexistence schemes has its own advantages and drawbacks based on the application served. For instance, the goal of the first scheme is the mitigation of mutual interference while guaranteeing satisfactory performance for both functions. Meanwhile, the second category avoids spectral overlap by enabling radar to sense the communication channel at very low rates. The last category joins the functionalities through hardware and no resource negotiation takes place. Hence, it is necessary to scrutinize these schemes and engineer \ac{THz}-tailored schemes can meet the particular needs of a given application.  
	\item \emph{Waveform design:} The designated waveforms for radar sensing are typically unmodulated single-carrier signals or short pulses and chirps. As such, this results in high power radiation and simple receiver processing \cite{rahman2020enabling}. In contrast, communication signals consist of a mix of unmodulated (pilots) and modulated signals. This results in a higher transceiver complexity. Henceforth, joint systems need to take into account these waveform differences and characterize tradeoffs in a way to jointly optimize the performance.
	\item \emph{Design complexity:} Extending joint sensing and communication to long-range use cases and ubiquitous coverage requires integrating \ac{THz} with \ac{mmWave}. Subsequently, implementing a dual-frequency band system complicates the feedback mechanism between sensing and communications. Also, a dual-frequency band system leads to problems in sensing scheduling schemes due to differences in the interference, accuracy level, and ranges of these frequency bands.
\end{itemize}
\indent Indeed, joint sensing and communication systems will have a central role for wireless \ac{THz} networks and next wireless generations. This results from the symbiotic integration of a high carrier frequency, immense bandwidth, quasi-optical characteristics, and natural use of \acp{RIS}. In fact, a significant use-case of the sensing feedback for communication takes place in the initial access stage. As such, sensing paves the way for the initiation of a successful channel estimation process. Hence, we next elaborate on the challenges and the transformative solutions needed at the PHY-layer for a successful channel estimation and initial access process in \ac{THz} wireless systems.
\section{PHY-layer Procedures}
\renewcommand{\arraystretch}{1.5}
\begin{table*}[t]
	\footnotesize
	\caption{Summary of the impact and challenges associated to each one of the THz-enabling solutions.}
	\centering
	\begin{tabular}{m{5.5cm}m{5.5cm}m{5.5cm}}
		\hline
		\textbf{THz Enabling Solution} & \textbf{\hspace{0.4cm}Impact} & \textbf{\hspace{0.4cm}Challenges} \\
		\hline
		\textbf{Cell-free massive MIMO}&\begin{itemize}
			\item Reduction of handovers and handover failures.
			\item Suppression of interference.
			\item Local CSI with lower overhead.
		\end{itemize}& \begin{itemize}
			\item Highly varying dynamic user-centric clustering.
			\item Connectivity of all \acp{SBS} to a cloud.
			\item Complex distributed network operation.
		\end{itemize}\\
		\hline
		\textbf{RIS-enhanced architecture}& \begin{itemize}
			\item Increased \ac{LoS} probability.
			\item Improved multi-path sensing.
			\item Cost efficiency.
			\item Generation of \acp{OAM} through meta-surfaces.
		\end{itemize} & \begin{itemize}
			\item Complexity in aggregating CSI.
			\item Strategic localization of RISs for enhanced cooperation.
		\end{itemize} \\ 
		\hline
		\textbf{Integration with mmWave/sub-6 GHz} & \begin{itemize}
			\item Upgraded \ac{THz} service to outdoor and highly mobile services.
			\item Enhanced reliability.
		\end{itemize} & \begin{itemize}
			\item Complexity in resource allocation between the three frequency bands.
			\item Abundance of THz ISs in contrast to the density of mmWave/sub-$\SI{6}{GHz}$ \acp{SBS} .
		\end{itemize}\\
		\hline
		\textbf{OAM} & \begin{itemize}
			\item Higher spectral efficiency.
			\item Novel multiplexing and multiple access dimension through OAM modes.
		\end{itemize}& \begin{itemize}
			\item Lack of wireless models to characterize OAM modes.
			\item Decoding data from OAM carrying modes successfully. 
			\item Compliance with existing wireless infrastructure. 
		\end{itemize}	\\
		\hline
		\textbf{NOMA} & \begin{itemize}
			\item Higher spectral efficiency.
			\item Efficient resource allocation to worst-case scenarios.
		\end{itemize} & \begin{itemize}
			\item Low received power.
			\item Substantial interference.
		\end{itemize} \\
		\hline
		\textbf{Joint sensing and communication systems} & \begin{itemize}
			\item Improved network learning by augmenting channel data with sensing data.
			\item \acp{UE} granted with 3D mapping and situational awareness capabilities.
			\item Accessibility to a broader range of applications that require radar/sensing capabilities.
		\end{itemize} & \begin{itemize}
			\item Dynamic spatial multiplexing amid contrasting beam requirements in communications and sensing.
			\item Complexity in coexistence approach and resource management.
		\end{itemize} \\
		\hline
		\textbf{Distributed multi-agent network optimization} & \begin{itemize}
			\item Various network patterns and service trends broken down into dynamic clusters.
			\item Hierarchical structure leveraged to generalize and specialize into network behavior/specific service requirements.  
		\end{itemize} & \begin{itemize}
			\item Seamless cooperation between multi-agents.
			\item Scarcity of big datasets.
			\item Intra-cluster variations due to time-varying data.
		\end{itemize} \\
		\hline
		\textbf{Meta-learning driven network optimization} & \begin{itemize}
			\item Mapping heterogeneity in behavior to multiple tasks.
			\item High generalizability.
		\end{itemize} & \begin{itemize}
			\item Difficulty in defining tasks vis-à-vis different requirements.
			\item Violation of HRLLC requirements and near real-time operation.
		\end{itemize} \\
		\hline
	\end{tabular}
	\label{table:Vision}
\end{table*}
PHY-layer procedures, such as channel estimation and initial access, face new challenges in \ac{THz} frequencies due to multiple reasons. First, \ac{THz} channels are highly dimensional and often very sparse due to their beam representation. Indeed, the high number of antenna elements results in a high number of pilots, leading to significant overhead. Second, the very narrow-beamed \ac{THz} links of the \acp{SBS} and their corresponding \acp{UE} cannot meet in space at initial access, i.e., prior to channel estimation and any information exchange. This results in the so-called \emph{deafness problem}. While this problem was encountered at \ac{mmWave} frequencies, the key difference at \ac{THz} bands is that quasi-omnidirectional antennas cannot be used due to the higher propagation losses \cite{xia2019expedited}. Third, the highly varying \ac{THz} channel has a very small coherence time that must accommodate the combined duration of uplink training, downlink payload data transmission, and the uplink payload data transmission. In fact, the network will not only experience a coherence time significantly shorter than the one at \ac{mmWave}, but, also, the \ac{THz} \ac{EM} wave is highly susceptible to multiple factors such as molecular absorption, blockage, and minute beam misalignment. Hence, this further hinders guaranteeing the aforementioned combined duration below the \ac{THz} coherence time. These intertwined challenges imply that conventional low frequency protocols and schemes used for channel estimation and initial access cannot be used to capture the distinct features of the \ac{THz} channel behavior.
\subsection{Channel Estimation}
\indent \ac{THz} networks are likely to be deployed in a fairly dense distributed architecture, empowered with \acp{RIS}. Having multiple \acp{RIS} cooperating in cell-less architectures makes it very difficult to capture the instantaneous \ac{CSI}. On the one hand, the dynamic user-centric clusters are highly time-varying. On the other hand, multiple \acp{RIS} are operating in different modes, i.e., as an \ac{SBS} or intelligent reflector. These factors lead to non-stationary channel behaviors and compound channels respectively. Predicting partial \ac{CSI} might be feasible in such conditions, nonetheless, partial \ac{CSI} amid the highly varying \ac{THz} channel leads to poor beam alignment, user association, and network optimization, ultimately yielding a highly unreliable communication link. Capturing and characterizing an instantaneous \ac{CSI} is an essential building block. Therefore, it is necessary to exploit the data used to build partial \ac{CSI} to predict accurately the instantaneous \ac{CSI} of the network.\\
\indent Every user-centric cluster exhibits unique channel conditions that are highly varying with time. \emph{First}, the limited number of \acp{UE} within a user-centric cluster and its small geographical area leads to a scarcity in the channel data. \emph{Second}, the user-centric clusters exhibit high correlations between each other that need to be capitalized. Hence, these factors require on the one hand supplementing the channel data to improve the generalizability of channel estimation. On the other hand, multiple agents need to be deployed to cooperatively exchange \emph{generalizable} channel behaviors across user-centric clusters, while specializing by the use cases exhibited within their cluster. Thus, this calls for novel multi-agent generative \ac{ML} mechanisms. In such mechanisms, each agent will generate synthetic environments from the available small data to train and estimate the channel. In fact, the concept of generative networks has started seeing light in channel estimation. For instance, in \cite{balevi2020high} the authors performed channel reconstruction by eliminating the need for a priori knowledge for the sparsifying basis and capturing the deep generative model as a prior. The work in \cite{kasgari2020experienced} proposed a \ac{GAN} approach that pre-trains a deep-\ac{RL} framework using a mix of real and synthetic data to assimilate a broad range of network conditions. Clearly, the approaches adopted in \cite{balevi2020high} and\cite{kasgari2020experienced} are still in their infancy given their dependence on a single learning agent and their inability to learn compound channel models for \ac{RIS} enhanced networks.  Thus, such techniques should be extended to handle the highly varying \ac{THz} channel as well as the need for cooperation among different learning agents and their corresponding user-centric clusters. Here, one can build on our recent results in \cite{qqzhang2020} and \cite{zhang2021distributed} which showed that the use of fully-distributed \acp{GAN} ~\cite{ferdowsi2020brainstorming} (i.e., generative models that go beyond classical federated learning) can be effective for \ac{mmWave} channel estimation in multi-drone networks.
\subsection{Initial Access}
\indent Having a successful initial access at \ac{THz} frequencies requires finding a solution that allows the beams of the \acp{SBS} or \acp{RIS} and their corresponding \acp{UE} to meet in space (prior to any information exchange). Hence, to address this so-called \emph{deafness problem}, one can envision two prospective solutions:
\begin{itemize}
\item Leverage the integration of \ac{THz} networks with lower frequency bands to perform the link configuration, beam association, and gather all the control information prior to information exchange
\item Explore \ac{THz} sensing and localization capabilities. In other words, instead of solely relying on communication data exchange, one can rely on sweeping sensing beams to acquire a \emph{situational awareness} prior to any information exchange. As such, this situational awareness continuously updates the \acp{SBS} with the location and orientation of the potential \acp{UE} they will associate to.
\end{itemize}  For the sensing solution, relying on a single narrow-beam \ac{LoS} link to obtain sensing data is highly time consuming and subject to errors due to frequent blockages. As such, to obtain richer information about the environment, \acp{RIS} with a high number of metasurfaces can create multiple independent paths characterizing richer information of human positions and orientations. In turn, this will successfully set the stage for a \emph{spatial rendez-vous} between the communicating beams prior to the channel estimation period. In fact, beam management for dynamic scenarios at \ac{THz} frequencies must be maintained \emph{instantaneously} given the high susceptibility of narrow beams to the fluctuations in user orientation. 
Thus, situational awareness can provide the needed instantaneous feedback, owing to the \ac{THz}'s highly accurate sensing capability as a byproduct of its large bandwidth. Next, after laying out the foundation needed in terms of architectures and systems of operation for versatile wireless \ac{THz} systems with joint sensing and communication functions, we will discuss the measures that allow us to improve the spectrum efficiency and support more users.
\section{Spectrum Access Techniques}
Conventional spectrum access schemes like \ac{OFDMA} and \ac{CDMA} used at lower frequency bands and previous wireless generations cannot be directly applied to \ac{THz} wireless systems. On the one hand, the peculiar quas-optical propagation features and its corresponding hardware constraints make it difficult to directly employ traditional spectrum access techniques \cite{wei2018multi}. On the other hand, the stringent requirements of emerging 6G services in terms of delay and resources require more efficient spectrum access techniques. For instance, while \ac{TDD} is a widely adopted multiplexing technique for current massive \ac{MIMO} systems, yet it still leads to an overhead latency that is added to the \ac{E2E} delay. This calls for alternative techniques that do not rely on time resources and that can eliminate such latency from the equation. Effectively, \ac{THz}'s quasi-opticality paves the way for many such techniques that do not rely on traditional time, frequency, and space resources. Hence, we next delve into these \emph{spectrum access \ac{THz}-tailored techniques} that lead to an improved multiplexing and multiple access efficiency.
\subsection{\Ac{OAM}}
The concept of \ac{OAM} \cite{ni2020electromagnetic} is a physical property of \ac{EM} waves that has recently drawn attention as means to dramatically  improve the channel capacity and the spectral efficiency of communication systems. Effectively, the roots of this physical property pertain to the rotation of optical beams. This rotation is characterized through an angular momentum that has two components: the first component is the polarization vector rotations called \emph{spin}, whereas the second component of substantially higher magnitude is the phase structure rotations called \emph{orbital angular momentum}. Moreover, to enable carrying information, the internal \ac{OAM} is a helical wavefront that enables boosting the optical and quantum information capacities through a spatial modal basis set \cite{yao2011orbital}. As such, the literature uses the term \ac{OAM} in short to denote the internal \ac{OAM} when considering it as an independent information carrier.\\
\indent Unlike the lower frequency bands that showcase a minimal improvement with the adoption of \ac{OAM} \cite{edfors2011orbital}, \ac{THz}'s quasi-opticality enables it with a robust \ac{OAM} capability. Such a capability can be leveraged to provide a novel dimension for multiple access, multiplexing, and increased spectral efficiency. Particularly, \ac{OAM} has a great number of \emph{modes}, i.e., topological charges, that are orthogonal to each other. On the one hand, \ac{OAM} modes can enable \ac{THz} \ac{LoS} links to bypass conventional spatial multiplexing and deploy \ac{OAM} multiplexing. This could be a promising way to improve system capacity and provide an alternative to orthogonal frequency division multiplexing\footnote{OFDM is very complex to implement at the \ac{THz} band due to the high peak-to-average power ratio.} (OFDM) by mitigating the interference of narrow-beam \ac{LoS} links. On the other hand, these \ac{OAM} modes provide a new dimension for user multiple access without exhausting the time or frequency resources. This, in turn, preserves the high bandwidth available at \ac{THz} in spite of traffic intensive services, and without acquiring additional system delays. Additionally, this allows us to also multiplex resources among the \ac{OAM} modes to further increase the spectral efficiency of \ac{THz} communications. Owing to \ac{OAM}'s multiplexing power, an open research area is to investigate whether sensing and communication functionalities can be multiplexed using \ac{OAM} modes.\\
\indent Moreover, \ac{OAM} can be generated using metasurfaces by regulating their phases. Interestingly, as we discussed in Section IV-B, \ac{THz} networks are likely to heavily rely on \acp{RIS} that are made of a discrete or continuous number of metasurfaces. As such, capitalizing on the existing network architecture and exploiting metasurfaces allows us to achieve a milestone in spectrum efficiency, resource allocation, multiplexing, and multiple access. Thus, it is crucial to adjust different \ac{MAC} protocols in a way to conform with \acp{SBS} and \ac{UE} adopting \ac{OAM} multiplexing and multiple access schemes, i.e., achieving orthogonality in a domain independent of space, time, and frequency. Moreover, new signal design and processing methods are needed to optimize the generation and transmission of \ac{THz} \ac{EM} waves that adopt \ac{OAM} to carry information over the \ac{OAM} modes.
	\begin{figure*} [t!]
	\begin{centering}
		\includegraphics[width=.8\textwidth]{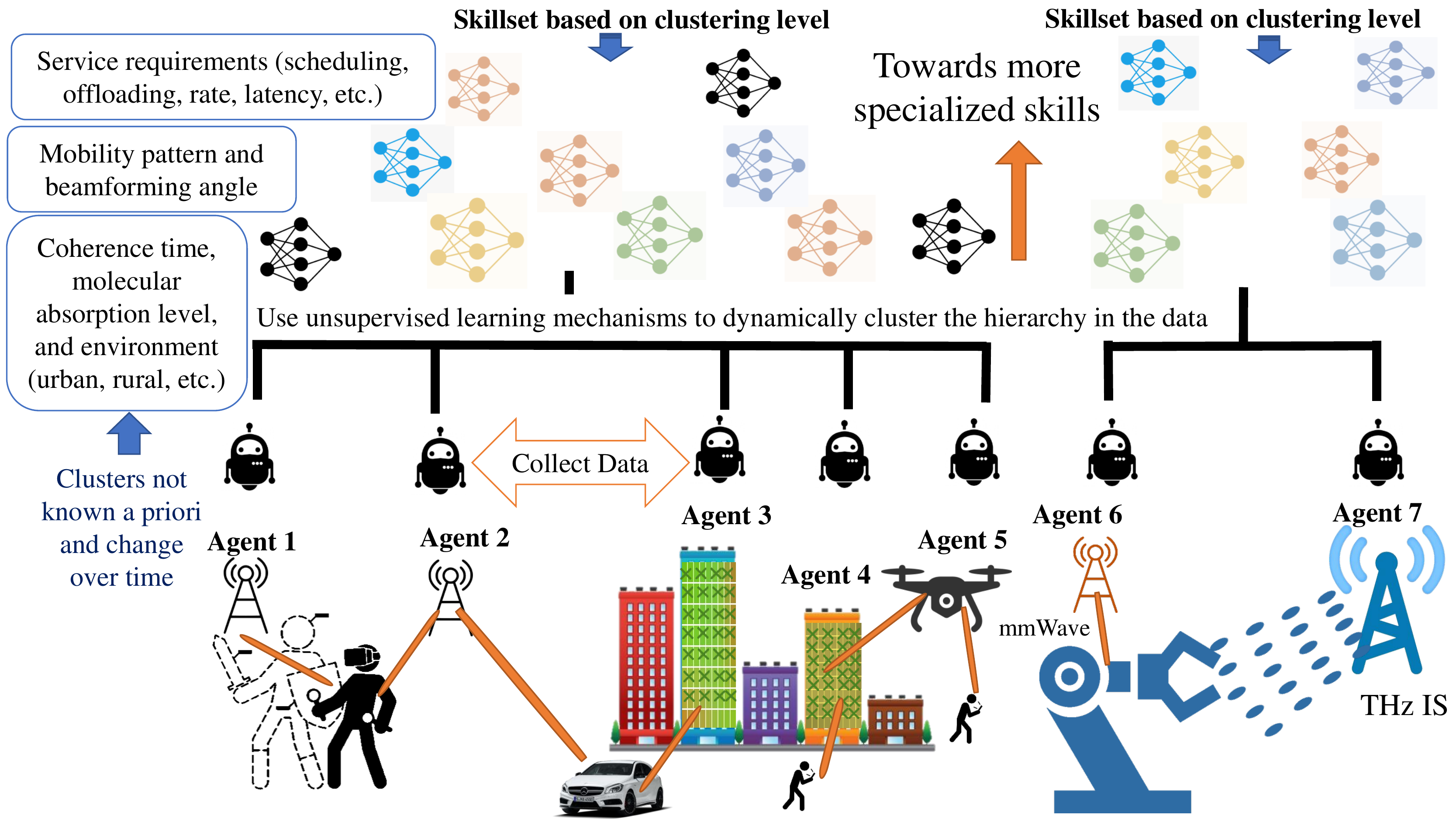}
		\caption{\small{Illustrative figure showing the need for generalized and specialized learning for 6G systems operating at THz bands.}}
		\label{fig:hierarchy]}
	\end{centering}
\end{figure*}
\subsection{\Ac{NOMA}}
While \ac{OAM} is a novel dimension that provides multiple degrees of freedom for \ac{THz} transmissions, in terms of multiple access and multiplexing, one can alternatively rely on \ac{NOMA} techniques. \ac{NOMA} is a multiple access technique that uses the power domain to introduce quasi-orthogonality. As such, it pairs up users experiencing higher channel gains with those facing lower ones, thus, reducing the disparity in the network by providing weak user greater power allocation \cite{naqvi2016combining}. Particularly to \ac{THz}, where the gap between the best-case user scenario and the worst-case user scenario is considerably large, \emph{\ac{NOMA} allows shrinking this gap and providing a fairer experience across users}. Henceforth, \ac{NOMA} can improve the service to users suffering from the dynamic and extreme network conditions such as deep fades, blockage, and user mobility. Such conditions traditionally lead to intermittent and unreliable \ac{THz} links, and, thus, \ac{NOMA} can provide better guarantees for the \ac{QoE} of critical services like holography and next-generation \ac{XR}.\\
\indent Furthermore, it has been shown that \ac{NOMA} performs better in systems that have high \ac{SINR} \cite{ding2014performance}. Given that \ac{THz} systems have considerably large \ac{SINR} due to the short communication range, adopting \ac{NOMA} schemes at \ac{THz} is highly beneficial and provides a higher spectral efficiency compared conventional orthogonal schemes. Nonetheless, due to the high directionality of \ac{THz} links and pencil beams, adopting single-beam \ac{NOMA} remains limited by the number of users that can be served. Furthermore, difficulties arise in providing suitable user pairing schemes with beamforming, as well as optimal power and bandwidth allocation. This calls for novel schemes that allow an efficient user pairing despite the high directionality of the beams \cite{zhang2019joint}. Moreover, in comparison to \ac{OMA}, \ac{NOMA} exhibits a lower received power and a more severe interference that must be properly managed for effective multiple access.\\
\indent To overcome the aforementioned challenges, one potential solution is to benefit from the \ac{THz} \ac{RIS} enhanced architectures. For instance, we can leverage the \ac{RIS}-\ac{NOMA} synergy through a network with \ac{THz} \acp{SBS} and \acp{RIS} performing passive beamforming. In that sense, relying on an \ac{RIS}-aided network improves the energy efficiency of the system. In turn, \ac{NOMA} provides \acp{RIS} architectures with a multiple access technique capable of enhancing massive connectivity and spectral efficiency. Hence, the \ac{RIS}-\ac{NOMA} synergy can be reaped whether \acp{RIS} are active or passive. Moreover, \emph{on the one hand}, \ac{RIS}-\ac{NOMA} based architectures, in contrast to massive \ac{MIMO}-\ac{NOMA} systems, can potentially overcome the fluctuations resulting from deep fades, blockages, and mobility by exploiting reflected links. \acp{RIS} overcome these events by providing
extended communication ranges and virtual \ac{LoS} links to \ac{THz} \ac{NOMA} systems. The ease of beam control through \acp{RIS} allows exploiting techniques like multi-beam \ac{NOMA} \cite{wei2018multi}. This concept is a beam splitting technique that relies on generating multiple beams to serve multiple \ac{NOMA} users over each radio frequency. Hence, it can mitigate the limitations resulting from narrow beams. \emph{On the other hand}, instead of conventionally determining the \ac{NOMA} decoding order of the users using their channel power gains, reconfiguring the \ac{RIS} phase shifts allows us to flexibly design the users' decoding order \cite{xiu20201reconfigurable}. Moreover, \acp{RIS} can improve the data rates of weak \acp{UE} without the need for additional transmit power \cite{de2020role}. Hence, it can provide an extra benefit without any additional energy consumption. 
Thus, \ac{RIS} networks introduce multiple degrees of freedom for improving the \ac{THz}-\ac{NOMA} performance.\\
\indent After equipping wireless \ac{THz} systems with effective multiple access and multiplexing schemes, fulfilling the diversified requirements and functions (sensing, communication, localization) of 6G services requires departing from conventional network optimization. Such methods are simply not adequate for \ac{THz} systems because of the real-time nature of the optimization needed on the one hand, and because of the highly varying \ac{THz} channel on the other hand. Thus, novel algorithmic approaches are needed so that \ac{THz} systems can ensure a real-time network optimization, these will be discussed next.
\section{Real-Time \ac{THz} Network Optimization}
\ac{THz} will mainly be the driver of high data rates and high-resolution sensing for 6G services. Effectively, \ac{THz} offers 6G applications many benefits in terms of abundant bandwidth, improved spectral efficiency, and enhanced localization. Nevertheless, its uncertain channel jeopardizes its robustness to provide real-time communication, control, and computing functionalities for these services. To enhance \ac{THz}'s robustness, the following caveats need to be taken into account:
\begin{itemize}
	\item \emph{Highly varying channel nature:} \ac{THz} has a highly varying channel, i.e., its coherence time is extremely short. At the other end, emerging 6G services like \ac{XR} for instance highly rely on a real-time response. In other words, the network needs to serve such applications \emph{continuously} without millisecond disruptions in the service. Henceforth, if the \ac{THz} beam tracking, resource allocation, and user association is performed based on an \emph{outdated} coherence time, the real-time communication will be disrupted.
	\item \emph{Susceptibility to extreme events:} \ac{THz}'s quasi-optical nature increased \ac{THz}'s susceptibility to molecular absorption, blockage, and minute beam misalignment. Such events inherently jeopardize the \emph{instantaneous} reliability of \ac{THz}'s links to serve 6G services.
	\item \emph{Heterogeneous frequency bands:} The coexistence of \ac{THz} with \ac{mmWave} and sub-6 GHz links (as outlined in Section V) calls for mechanisms capable of estimating the channel parameters, managing resources, and tracking beams despite the heterogeneous characteristics of communication and sensing over different frequency bands. 
	\item \emph{Joint resolution and rate optimization:} Deploying joint sensing and communication \ac{THz} systems requires the joint optimization of different objectives (e.g. high data-rates and high-resolution). Meanwhile, such objectives potentially dictate conflicting modes of operation. For example, beams used for communication must be narrow and sharply pointing, whereas sensing beams must be varying directional scanning beams. Henceforth, the real-time network control needs to be also assessed in terms of sensing.
	\item \emph{Compound channels:} \ac{RIS} enabled architectures mitigate multiple challenges pertaining to the \ac{THz} channel. Nonetheless, such an architecture leads to compound channels and an increased complexity whereby more than one link need to be continuously synchronized.
\end{itemize}
Given the aforementioned caveats and the lack of explicit models that allow us to clearly draw performance tradeoffs in terms of rate, reliability, latency, and synchronization.  One could exploit the concept of data-driven networks, and let the data collected in terms of channel measurements, sensing measurements, and \ac{QoS} measurements be the decisive factor to examine system performance and improve it. Nonetheless the following key challenges need to be examined:
\begin{itemize}
\item \emph{Non-stationary data:} The \ac{THz} channel and the key performance indicators of the \ac{THz} network such as handover, beam-tracking, and molecular absorption have time-varying and non-stationary distributions that are jointly correlated. Thus, predicting and generalizing these distribution patterns is inherently complex. Hence, this calls for mechanisms capable of breaking the correlation between the events in order to simplify the prediction process. 
\item \emph{Failure of centralized methods:} On the one hand, satisfying the low latency requirements of 6G cannot be met by wasting communication resources to access a centralized server. On the other hand, the distribution patterns predicted at a specific location in the \ac{THz} network (e.g. scheduling policy or cell-association) can be invalid and outdated at another location. Consequently, a single decision maker cannot generalize the \ac{THz} performance at different instances. Thus, this calls for learning frameworks that deploy multiple edge agents which can collect and locally learn the data.  
\item \emph{Scarcity of data:} Solely relying on location-specific data to learn distributions characterizing the \ac{THz} network performance will be challenging due to the insufficient training periods. Consequently, \ac{ML} algorithms would learn corner cases instead of \emph{generalizing} the distributions learned. To address this challenge, one should consider complementing existing channel data from other synthetic or real data to achieve improved training processes. Another alternative approach is to leverage the concept of theory-guided data science \cite{karpatne2017theory}. This concept integrates theoretical channel models with data science models to improve the scientific consistency of data science models.
\item \emph{Real-time response:} Current \ac{ML} methods still incur long training periods that will not allow learning agents to operate in real-time. Performing the training offline might not be a valid solution due to the non-stationarity of the data. In other words, the distributions learned offline are not valid when used to perform decision-making in an online fashion. Thus, this calls for \emph{real-time \ac{ML} methods} that can incur shorter training periods. 
\end{itemize}
Based on these key challenges, we next underline the need for a multi-agent learning framework capable predicting the generalized \ac{THz} network performance, while capturing peculiar specialties based on the specific use cases and events foreseen. 
\subsection{Towards Generalizeable and Specialized Learning} Addressing the intertwined and non-stationary traits of data in a \ac{THz} wireless system can be performed by recognizing the latent generalizable traits common among the resources and services being optimized. Subsequently, leveraging multi-agent learning in contrast to centralized \ac{ML} techniques allows us to extract the specific and specialized characteristics within a type of service, a mobility pattern, or a subset of \acp{UE} and resources. After the learning agents (e.g. an active \ac{SBS} or \ac{RIS}) have collected the data, dynamic clustering of the data can be performed by using unsupervised learning schemes. This clustering allows breaking the complex and joint correlations among different resources and data points, it also assigns each learning agent particular specialties based on the frequency of events seen in the data. For instance, multiple learning agents attempting to predict the mobility distributions of users exhibiting homogeneous mobility patterns will share the models learned. Effectively, this allows aggregating more data to achieve more robust training vis-a-vis the non-stationary data. Similarly, the same learning agents might potentially have different specialities pertaining to the scheduling policy adopted. These models are learned seperately by each agent to reduce the overhead and incremental delays. For example, one agent can be specialized in \ac{AR} use cases, low and medium mobility patterns, and a molecular absorption level that corresponds for indoor environments. Meanwhile a second agent that collaborates with this agent can also be specialized in \ac{AR} use cases, however, the skillsets acquired for mobility and molecular absorption correspond to highly mobile \acp{UE} and outdoor environments. Subsequently, the collaboration between such agents serves to strengthen the common skillsets, and to minimize communication resources on the exclusive skillsets learned.  \\ 
\indent Hence, one agent can be specialized in one or more skillsets, this depends on the level of heterogeneity in its channel data. Effectively, if a skillset is common among a high number of agents, the prediction capability of all these agents is improved vis-à-vis this skill. Furthermore, the learning capability of an isolated skillset (attributed to a single agent only) depends on the complexity of this skill and the amount of data gathered.  As such, our suggested \ac{ML} framework allows learning agents to benefit from the common denominator and shared specialties they exhibit, while reducing the overhead for heterogeneous patterns learned separately. The overall approach that we propose here is captured in Fig~\ref{fig:hierarchy]}. In this figure, we show an example in which dynamic clusters are formed based on three hierarchical levels. By moving to the top of the figure, the clusters become more specialized. Hence, this allow us to conquer and divide the intertwined trends and enable agents with a high generalizability and specialization.\\
\indent A recent distributed learning framework, dubbed \emph{democratized learning}, was proposed in \cite{nguyen2020self} and \cite{nguyen2020distributed}, in order to capture specialization and generalization in a network of learning agents. This framework is a potentially promising solution for the considered \ac{THz} wireless system problems because it provides means to mimic human cognitive capabilities by collaboratively performing multiple complex learning tasks. In this framework, the agents according to their different characteristics form appropriate groups that are tailored for a specific specialization. Such groups are self-organized in a hierarchical structure where the biggest group shares the most common knowledge across all agents, then groups start to shrink in size with more specialized skills. Nonetheless, in those prior works, such algorithms have only been applied on MNIST and Fashion MNIST datasets. As such, this framework was not tested in time-sensitive, scarce, and heterogeneous data from \ac{THz} networks. Also, these existing learning frameworks were primarily designed for simple classification tasks in contrast to the real-time reinforcement learning needed to control a wireless network. Thus, it is necessary to empower each learning agent with an engine capable of discerning the complex patterns in the data in a real-time fashion. In other words, the algorithms trying to predict and optimize the network process need to be empowered with more \emph{expressive power}.\footnote{Expressive power is the ability to represent a large number of learning algorithms, i.e., more expressive power means that the technqiue allows us to represent more sophisticated learning procedures \cite{finn2018learning}.} As such, to build intelligent wireless systems that can learn with the same versatility and flexibility as the human brain, one could exploit the concepts of multi-task and meta-learning which will be discussed next.
\subsection{Towards Multi-Task Learning and Meta-Learning}
While multi-agent learning allows breaking trends into clusters, the highly varying data structure of \ac{THz} systems leads to intra-cluster inconsistency. For example, even after clustering a group of \acp{UE} onto a common service type and a single mobility pattern, standard \ac{ML} methods like deep Q-learning might not be able to find reasonable solutions when faced with data that is slightly out of the distribution learned. In fact, this aspect is highly relevant for \ac{THz} systems due to their heavy tail as a result of their susceptibility to extreme events like blockages or deep fades. To mitigate these challenges, one could divide every single learning task problem into multiple tasks.\\
\indent For example, one could partition a typical \ac{THz} beam alignment problem into multiple learning tasks. It is important to note here that the term \emph{learning task} holds a  different significance than its semantic meaning in the English language. For instance, the presence of a high density of blockages and a low likelihood of \ac{LoS} might constitute one beam alignment task. Meanwhile, an average density of blockages consequently leading to a \emph{nominal \ac{THz} beam alignment process} is another beam-alignment task, i.e., one that would only take place under average environmental figures. \emph{In the first case}, the learning agent will typically  perform a risk-averse action, whereby the reward and the learning setting are tuned to account for a high number of extreme and catastrophic events. In that sense, for this first task, the learning agent's goal would be to guarantee a higher number of active links connected to a given user (this can be achieved, for example, by generating more independent links from active \acp{RIS} towards a \ac{UE}, or by optimizing more reflected links from passive \acp{RIS} towards a \ac{UE}). Hence, under the first task, for a given \ac{UE}, an exhaustive number of active links would be associated by the learning agent in order to guarantee at least a single perfectly aligned \ac{LoS} link. Also, because of this riskiness associated to the environment, the learning agent may continuously exhaust a high number of frequency, energy, and space resources to sensing functionalities. This is because of the intrinsic need to continuously process the user's instantaneous location and orientation in such an extreme environment. Hence, a significant network overhead would potentially incur to account for \acp{UE} under the first task, this is a result of the higher number of resources associated to sensing functionalities.\\
\indent \emph{In contrast, the second learning task}, that focuses on nominal operation, requires a lower level of caution by the agent. Here, the learning agent can allocate more resources to multiple \acp{UE} communication links (rather than possibly needing to exhaust wireless resources on sensing a single \ac{UE} under a risky environment) which naturally reduces the overhead incurred from the continuous feedback needed from \ac{SLAM} in the first task. In other words, instead of expending a significant amount of radio and energy resources to establish a high number of active communication links as well as a continuous sensing feedback for a small subset of \acp{UE} experiencing dire conditions, the decisions made by the agent in this second learning task can achieve a higher level of spectral and energy efficiency. As such, assimilating both learning tasks allows the learning agent to have a full view of the beam-alignment distribution in a \ac{THz} network. Thus, this contributes to a decision making process that can \emph{cautiously solve problems without being too restrictive}. Instead, if one is to approach this problem as a single task problem, then each learning agent will tend to learn an \emph{average} behavior vis-à-vis its experienced environment. In essence, this will not allow the agents to tune their cautiousness level according to the type of environment encountered. Also, note that, while the \emph{riskiness of the environment} and the blockage density are the varying parameter across tasks in this this illustrative beam alignment problem, another wireless problem will have a set of different parameters that could distinguish a task from another. \\
\indent  It is important to note that every learning problem exhibits a different number or type of learning tasks. In fact, often times, such tasks cannot be known a priori. Effectively, a single \emph{task} distribution is characterized by a collection of homogeneous data that typify it. Learning multiple tasks simultaneously allows us to capture an umbrella distribution that covers all the possible events. On the one hand, this enhances the expressive power and scalability of the learning agent vis-à-vis beam alignment in contrast to learning corner-cases. On the other hand, despite the scarcity of data and its inability to cover the full distribution of beam alignment, the learning agent can generalize for out of distribution events.\\
\indent Henceforth, the concept of multi-task and meta-learning can be leveraged  \cite{finn2017model} as an effective approach to address data scarcity and improve the generalizeability of a learning agent that operates in a complex environment. Nonetheless, applying such approaches to wireless \ac{THz} networks faces many challenges. The number and type of tasks that surround a particular networking problem are not always known a priori, and, thus, precisely defining a task is a key challenge. Furthermore, existing meta-learning techniques have been mostly procured on supervised learning problems \cite{nichol2018first, finn2017model,rajeswaran2019meta}, where data is inherently labeled. In contrast, wireless networking problems are effectively stochastic \ac{RL} problems in which a learning agent is fed with \emph{data environments}. Dealing with such data environments is cumbersome which is why \ac{RL} algorithms tend to be traditionally slow when applied to stochastic problems that have large state and action spaces. In essence, such algorithms have very long training periods, whereas 6G services need to operate in a low latency realm. Thus, such long training periods could disrupt the real-time operation of these services. Reducing the \ac{RL} training and exploration periods can be done through many measures. First, meta-\ac{RL} solutions potentially reduce training periods because they transform datasets to task specific datasets and, thus, they acquire fast adaptation to dynamic and potentially non-stationary environments. However, meta-\ac{RL} algorithms may be still incapable of continuously building upon previous experiences in a way to reach near real-time training periods through generalizability. Second, to further reduce exploration and training periods, the \ac{RL} structure needs to be re-architected whereby the data environments are processed or complemented with synthetic data before being directly used. Third, given the lack of labels in an \ac{RL} setting, designing the reward function is crucial in the convergence of the algorithm. Additionally, the uncertain \ac{THz} channel makes it difficult to explicitly define an oracle reward function for every single problem encountered. One method to overcome this challenge is through inferring reward functions using inverse meta-\ac{RL} \cite{xu2019learning}.\\
\indent Here, we note that some recent works such as \cite{maggi2020bayesian} proposed the use of Bayesian optimization with Gaussian processes as a potentially superior alternative to \ac{RL} (in terms of convergence and interpretability), for solving radio resource management problems. However, using this technique for real-world wireless problems is prohibitive because it requires satisfying multiple conditions such as a low number of control parameters, a smooth performance function, and a low update frequency to cope with the environmental dynamics. In fact, many of the functions dealt with in wireless problems, particularly with \ac{THz} systems are not smooth (e.g. because of factors such as molecular absorption). Also, 6G services mandate a near real-time network control, thus necessitating a high update frequency. For these reasons, it is natural to posit that \ac{RL}'s universal and flexible setup still constitutes the fundamental building block necessary for network optimization and control and, as such, novel \ac{ML} methods will have build on its basis. In that sense, adopting meta-\ac{RL} can be one approach that scales up \ac{RL} for wireless and non-wireless settings. Meanwhile, meta-\ac{RL} is still in its nascent stage and is an open research area which is ripe for exploration in the realm of 6G.  \\ 
\indent Now that we have provided a detailed panorama of the seven defining features of \ac{THz} wireless system, an overview of the \ac{THz} enabling approaches is summarized in Table~\ref{table:Vision}. Next, we discuss the most prominent \ac{THz} use cases that can potentially leverage those seven unique features.
	\begin{figure*} [t!]
	\vspace{-0.3cm}
	\begin{centering}
		\includegraphics[width=.8\textwidth]{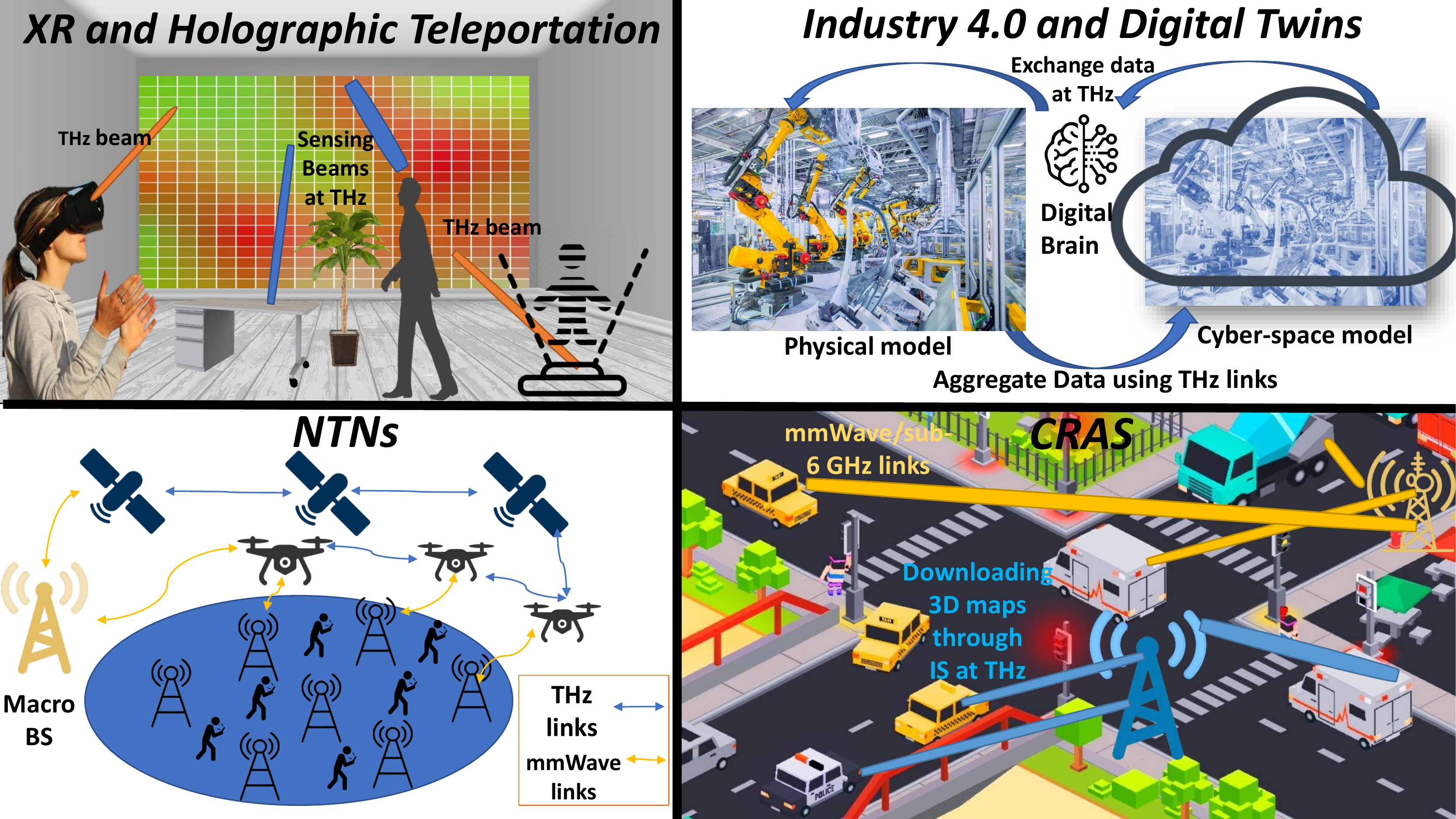}
		\caption{\small{Illustrative figure showcasing the four main uses cases for \ac{THz} systems.}}
		\label{fig:apps}
	\end{centering}
\end{figure*}
\section{Use Cases for \ac{THz} Wireless Systems} 
Fig.~\ref{fig:apps} presents the major 6G use cases that can potentially adopt \ac{THz} frequency band and exploit its seven defining features. Based on their distinct needs and mode of operation, these use cases can be deployed on different potential \ac{THz} architectures.
\subsection{XR and Holographic Teleportation}
\ac{XR} encompasses \ac{AR}, \ac{MR}, and \ac{VR}. \ac{VR} services will immerse the user in a seamless experience, while \ac{AR} services will overlay virtual components to the user's real time experience. Such applications have stringent \ac{HRLLC} requirements \cite{chaccour2020can}, because of the need to maintain the joint quality of visual and haptic components, and the need to sustain a high \ac{QoE} to the user. Furthermore, the definition of \ac{MR} is still not very concrete in the literature. Nonetheless, \ac{MR}'s main objective is to combine the capability of both technologies $-$ \ac{AR} and \ac{VR} in the same device \cite{speicher2019mixed}. With the evolution of 3D imaging, \ac{MR} is morphing to create the highly coveted \emph{holographic teleportation} application domain. In addition to their more stringent \ac{HRLLC} requirements, i.e., a rate of $\SI{5}{Tbps}$ \cite{li2018towards}, holographic flows require very tight synchronization in terms of feedback of five senses.\\
\indent Clearly, only \ac{THz} communications can cater to the potential of \ac{XR} and holographic teleportation by delivering the extremely high rates needed. Nonetheless, satisfying the rate requirement does not necessarily lead to a good \ac{QoE}. In fact, \ac{XR} and holographic services have diversified requirements that include, along with the high-rate needs, continuous low latency and jitter requirements, a need for fresh information (particularly for \ac{AR}), and synchronization among all five senses in holography. Furthermore, these requirements are not only diverse, but they also need to be satisfied with a high precision and accuracy. For instance, a momentarily disruption in the \ac{THz} \ac{LoS} link will lead to a disrupted experience. Thus, it is necessary \emph{first} to deploy \ac{THz} nodes on an architecture that increases the likelihood of \ac{LoS} links in indoor areas such as active and passive \acp{RIS}, while increasing the cost-efficiency of the network \cite{chaccour2020risk}. \emph{Second}, to guarantee a seamless \ac{XR} and holographic experience, different types of sensing functionalities need to be leveraged in a way to: a) Provide high precision and high resolution information about subtle changes in the user orientation and movement, which can be performed by exploiting the high-resolution \ac{THz} sensing feedback, b) Surround the user with a situational awareness of short range blockages, which can be performed by enabling multiple independent \ac{THz} sensing path with the use of \acp{RIS}, and c) Take the situational awareness one step further to account for farther surrounding objects, which can be performed by \ac{mmWave} sensing. \emph{Third}, the aforementioned sensing data and network communication data should be intelligently processed and then fed to the real-time \ac{ML} algorithms described in Section IX. This leads to a cross-layer intelligent system capable of overcoming \ac{THz}'s uncertainty to satisfy a plethora of requirements thus guaranteeing a high user \ac{QoE}.\\
\indent Extending such services to larger scale outdoor scenarios faces multiple challenges. On the one hand, the molecular absorption and the longer communication ranges significantly jeopardize the achievable rates by \ac{THz}. On the other hand, relying on \ac{THz} \acp{IS} in a predominantly \ac{mmWave}/sub-6 GHz \acp{SBS} networks cannot guarantee the instantaneous rates needed by \ac{XR} and holographic services. This leads to a tradeoff between the versatility of operation and the user \ac{QoE}. One prospective opportunity is to explore foveated rendering and compression techniques \cite{kaplanyan2019deepfovea} capable of providing savings in \ac{XR} content size. Furthermore, outdoor services tend to exhibit higher mobility use cases (e.g. \ac{AR} for assisted driving), such use cases have their particular challenges and open problems: a) outdated information in such use-cases leads to hazardous damages, thus, the freshness of the uplink here is highly significant, b) the increased mobility increases the overhead needed for beam tracking and mobility management, and c) a \emph{safe- seamless user experience} tradeoff takes place, optimizing such a tradeoff heavily relies on the techniques adopted to hybridize the network using lower frequency bands (as outlined in Section V).
\renewcommand{\arraystretch}{1.5}
\begin{table*}[t]
	\footnotesize
	\caption{Characteristics contrasting the needs of different 6G services.}
	\centering	
	\begin{tabular}{ m{3cm}  m{3.5cm}  m{3.55cm}  m{2.9cm}  m{3.2cm}}
		\hline
		\textbf{Key metric} & \textbf{XR and Holography} & \textbf{Industry 4.0 and Digital Twins} &\textbf{CRAS} &\textbf{\acp{NTN}}\\
		\hline
		\textbf{Potential mode of operation}  & \begin{itemize}
			\item Indoor or confined spaces
			\item Standalone \ac{THz} architecture (mmWave cannot satisfy the rate requirements)
		\end{itemize}& \begin{itemize}
			\item Controlled setting
			\item Standalone \ac{THz} architecture 
		\end{itemize} & \begin{itemize}
			\item Outdoor
			\item THz \acp{IS}
		\end{itemize} & \begin{itemize}
			\item Integrated fronthaul and backhaul
			\item mmWave/THz dual-band architecture 
		\end{itemize} \\
		\hline
		\textbf{Mobility support}  & Low/Medium & Low & High & Medium \\ 
		\hline
		\textbf{Rate}  & Order of $\SI{}{Tbps}$ & Depends on update between cyber and physical twins & Order of  $\SI{}{Tbps}$  & $\geqq \SI{1}{Tbps}$ \\ 
		\hline
		\textbf{Latency}  & \begin{itemize}
			\item $\SI{5}{ms}$ to achieve motion to photon latency
			\item Sub-millisecond for haptic capabilities
		\end{itemize}& $ 0.1-\SI{1}{ms}$ round trip time&  $\SI{1}{ms}$ round trip for reaction time & Application dependent \\ 
		\hline
		\textbf{Reliability paradigm}  & High downlink reliability and five senses synchronization & High bidirectional reliability & High bidirectional reliability & High connection density \\ 
		\hline
		\textbf{Service characteristic} & Transmitting real-time multi-sensory experiences & Cyber twin mimics the physical twin (especially in mission critical scenarios) & Real-time high definition content, in optimal time-space boundaries & Manage a plethora of aerial platforms such as \acp{UAV} and satellites \\
		\hline
		\textbf{Major challenge} & Tracking the micro-orientation and micro-mobility of users to maintain reliability & Synchronizing different building blocks of the complete digital twin & \begin{itemize}
			\item Strategically deploy \ac{THz} \acp{IS}
			\item Risk-aware autonomy
		\end{itemize} & 3D coverage\\
		\hline
		\textbf{Use of sensing} & Guiding every communication link with 3D \ac{SLAM} input and situational awareness of the surrounding environment & Sensing environmental changes in the physical-space with risky outcomes (e.g. disoriented controller) & Equipping \ac{CRAS} with multi-range and multi-resolution radar capabilities & Providing information about the Doppler effect and increased speeds in air and space. \\
		\hline
		\textbf{Opportunity that favors THz} & Indoor settings limited by shorter range of communication & Setting is controlled and not subject to high mobility & \acp{IS} are assisted by mmWave/sub-6GHz & Molecular absorption and losses are lower at heights above $\SI{16}{km}$
	\end{tabular}
	\label{table:services}
\end{table*}
\subsection{Industry 4.0 and Digital Twins}
The evolution towards Industry 4.0 is leading towards highly autonomous operations among machines and robots, requiring only occasional human intervention. In light of this, high precision and high accuracy manufacturing processes require novel instantaneous control mechanisms. Particularly, the rapid development of such systems and their automated processes requires data rates in the order of $\SI{}{Tbps}$, a latency in the order of hundreds of microseconds, and a connection density of $10^7/\text{km}^2$ \cite{giordani2020toward}. Meeting such high data rates can be naturally performed by deploying \ac{THz} networks. However, the use of \ac{THz} networks for Industry 4.0 applications brings forth many unique challenges that must be overcome in order to achieve near-zero latency, dense coverage, and precision-driven control mechanisms.\\
\indent Moreover, present industrial systems require the collection of large volumes of data from different sensors. Nonetheless, such data is only locally available and limits the flexibility of designing, developing, and preventing unwanted situations in large-scale autonomous systems. To provide an \ac{E2E} digitization, the concept of \emph{digital twins} \cite{rasheed2020digital} and \cite{farsi2020digital} has recently emerged as a means for creating a model for complex physical assets, thus scaling up the digitization of complex industrial structures and empowering them with full autonomy. Such digital twins should be characterized with trustworthiness so that engineers can rely on the remote control of physical systems by manipulating its cyber-space counter model.\\ 
\indent Providing such high-fidelity representation of the operational dynamics puts a burden on the underlying wireless network. In particular, the physical counterpart can only be fully mimicked by enabling a real-time synchronization between the cyber-space and physical spaces \cite{lu2020digital}. Given the large number of sensors continuously aggregating data to update the cyber-space model, the connection needs to be delivered at extremely high data rates at \ac{THz} frequencies. In turn, this synchronization imposes stricter bi-directional reliability requirements across thousands of devices with near-zero response times. There are several open problems in this area. First, one must investigate whether \ac{THz} systems can deliver bi-directional reliability for such a large amount of devices and real-time data transfer translated by very low delay jitter (in the order of $\SI{1}{\mu s}$). Subsequently, in case they fail \ac{mmWave} alone cannot satisfy the rate requirements, and thus, it is necessary to examine whether dual-band \ac{mmWave}/\ac{THz} systems can cooperate to update the cyber-space model in an \ac{HRLLC} fashion. Furthermore, extending the \ac{THz} coverage further calls for exploring some of the multiplexing and multiple access schemes discussed in Section VII. Such schemes open the door for multiple opportunities like improving the scalability of digital twins, accounting for a higher number of devices replicated, and improving the spectral efficiency.\\
\indent Digital twins not only allow real-time control, but they are also used to make predictions on the evolutionary dynamics and future states of large scale systems. This in turn enables anticipating failures, optimizing the system to design novel
features and to guide the decision making process. Thus, offloading data from physical models is highly error-sensitive, especially in the initialization of new processes. Faulty initialized cyber-space models will have biases that will propagate throughout the whole cycle of this process. Henceforth, communication links need to be driven by novel \ac{ML} models that combine real-time \emph{small data} and  control theoretic models \cite{karpatne2017theory} to build cyberspace models with higher accuracy. However, this process incurs large overhead to the transmitted content. Here, novel \ac{THz} control scheduling schemes need to be investigated. As a matter of fact, providing digital twins with real-time control amid their uncertain industrial setting can be performed by using \ac{ML} algorithms adopted in real-time \ac{THz} network optimization (the extreme events of the \ac{THz} channel and extreme events of industrial settings share a common denominator). In other words, \ac{ML} techniques adopted for the real-time optimization of \ac{THz} networks (elaborated in Section IX) are not limited to wireless networks, but can be later passed on to different real-time control environments. Subsequently, minimizing the prediction processing latency is the biggest delaying factor of digital twins, given that the cyber-space and physical-space are always mutually communicating using high \ac{THz} data rates.  
\subsection{CRAS}
\vspace{-0.1cm}
\ac{CRAS} services include autonomous driving, autonomous drone swarms, and vehicle platoons, among others. To be driven by full autonomy, such systems need to exchange large amounts of data such as high-resolution real-time maps, with their environment, e.g., other vehicles, or \acp{BS}. Additionally, such cyber-physical systems are often characterized by a high mobility and need to accurately sense and track their environment so as to determine their route optimization, and traffic and safety information \cite{zeng2019joint}. Thus, such systems require simultaneous sensing and communication for short range and medium range communications. Furthermore, such systems not only consume a huge amount of data to maintain their autonomy, but they also generate large volumes of data that could be of different types (e.g. 3D video of road conditions, radar data from nearby vehicles and objects). Henceforth, the wireless system should provide bidirectional high rates and reliable communication at the uplink and the downlink. As such, \ac{THz} frequencies can play an important role in enhancing the bidirectional rate for \ac{CRAS}. \\
\indent Although \ac{THz} systems can provide the rate requirements needed for CRAS, given the high mobility of \ac{CRAS} devices, the system reliability will be disrupted due to intermittent links and the unavailability of continuous \ac{LoS} links. Henceforth, to mitigate this challenge, \ac{THz} \acp{IS} can provide the high rates needed for high-resolution real-time maps, while being complemented by \ac{mmWave} and sub-6 GHz links to exchange less data intensive content, as indicated in Section V. To deploy CRAS over THz networks, several key challenges must be addressed. For instance, characterizing the \ac{THz} propagation in different outdoor environments (e.g. highway, urban, etc.) is an important challenge because of the high variability of \ac{THz} propagation. Moreover, one must develop new approaches for optimizing the location and density of \ac{THz} \acp{IS} versus \ac{mmWave} and sub-6 GHz links. Another key challenge is to provide an energy-efficient coverage despite the growing \acp{SBS} density. Furthermore, \ac{CRAS} can benefit from joint sensing and communication configurations. Particularly, given its longer range of communication, \ac{CRAS} can utilize \ac{mmWave}'s radar capability to detect objects and major environmental changes. Meanwhile, it can also exploit \emph{\ac{THz}'s high resolution sensing} to track subtle-moving targets and micro-mobility changes. The collected sensing data can augment the communications measurements to provide predictive control driven by \ac{HRLLC}. Hence, such systems will witness a synergy of integrated \ac{THz} and \ac{mmWave} sensing and communication. This synergy further calls for novel network modeling schemes, increased coverage to account for more devices, and novel predictive resource management schemes.\\
\indent Furthermore, \ac{ML} mechanisms controlling the autonomy of these systems need to act and learn reliably in the presence of out of distribution events \cite{filos2020can} (for example, collision avoidance systems in autonomous vehicles need to account for sudden extreme events like a sudden pedestrian crossing the street). Adopting  centralized black-box machine learning models here fails to provide strategic decision learning mechanisms, capable of  acquiring  and  accounting  for  the  uncertainty  in  a human methodical manner. For instance, such \ac{ML} methods might learn spurious relationships that might lead to misleading interpretations. The paucity of datasets exacerbates such interpretations further and can potentially lead to hazardous damages in \emph{high-risk settings} like vehicular environments. Thus, such systems need to be driven by novel trustworthy and real-time \ac{ML} mechanisms. To act instantaneously, in contrast to centralized \ac{ML}, multi-agent \ac{RL} mechanisms can be leveraged whereby agents perform local decision making without consuming extra computing and communication resources. Furthermore, to improve the generalizability of the learning agents, such agents can use \ac{HF} bands such as \ac{THz} and \ac{mmWave} frequencies to share their local models and/or data. Adopting such distributed schemes can reduce the bidirectional overhead in centralized \ac{ML} mechanisms and allow for a better cooperation as pointed out in Section IX. Clearly, the success of \ac{CRAS} is contingent upon developing a framework for providing autonomous control for such systems through wireless \ac{THz} systems. This framework will likely be characterized by explainable and low latency intelligence, high resolution sensing, and \ac{HRLLC} bidirectional communications. 
\subsection{\acp{NTN}}
\vspace{-0.1cm}
6G systems are expected to be characterized by ubiquitous 3D coverage, which can be provided by integrated space-air-ground communications. In fact, with the emergence of 5G systems, 3GPP started initiating plans for supporting \ac{NTN} to provide wide coverage and improve scalability \cite{anttonen20193gpp}. Effectively, by 2020-2025, more than 100 geostationary earth orbit and mega-constellations of \ac{LEO} based high throughput satellite systems with the capacity of Tbps will be launched \cite{giambene2018satellite}. On the one hand, \emph{compared to lower frequency bands}, \ac{THz} frequencies can provide air-to-air communications at extremely high data rates owing to its ultra high bandwidth. Furthermore, the attenuation of \ac{THz} links and their inability to penetrate the troposphere eliminates terrestrial spectral noise, interference, and jamming \cite{mehdi2018thz}. On the other hand, \emph{compared to optical communications}, \ac{THz} links have considerably larger beamwidths, thus, making the beam positioning and alignment more practical. In fact, fast and precise electronic beam alignment is possible when adopting phase-array antenna architectures which are a unique feature of \ac{RF} frequencies \cite{nagatsuma2018terahertz}. Additionally, \ac{THz} links do not suffer from molecular absorption in higher altitudes and free space, this opens the door for longer communication ranges. These advantages render the deployment of \ac{THz} frequencies for air-to-air links a natural choice.\\
\indent The main goals of non-terrestrial networks in 6G will be: a) Ensuring service continuity for highly mobile platforms (e.g. airplanes, trains) , b) Providing ubiquitous access by reaching out to under-served areas, and c) Enhancing the network scalability and providing more efficient backhaul. In light of these goals, space aerial communications are envisioned to migrate towards a new class of integrated \acp{UAV} and miniaturized satellites consisting of \ac{LEO} satellites and CubeSats. This migration is a result of their low costs associated with their production and deployment. Subsequently, such satellites have lower power transmission capacities compared to conventional satellite specifications. Furthermore, as shown in Fig.~\ref{fig:apps}, the use of \ac{THz} air-to-air links allow providing higher capacity backhaul links than fiber-optics. Subsequently, air-to-ground's larger communication ranges necessitate \ac{mmWave} links. Thus, the successful synergy of \ac{THz} and \ac{mmWave} links in air-to-air and air-to-ground links will allow providing a continuous links, ubiquitous communication, and high capacity backhauls.\\     
\indent While the use of \ac{THz} and \ac{mmWave} links in \acp{NTN} paves the way for multiple opportunities, several challenges and open problems need to be considered. \emph{First}, integrated satellite-terrestrial backhaul networks utilizing \ac{THz} and \ac{mmWave} bands must co-exist with current satellite and terrestrial systems. For instance, such a deployment will lead to interference from terrestrial backhauling transmitters to the satellite backhauling terminals. Thus, the development of flexible spectrum sharing techniques is needed to maintain suitable isolation between different network operations. \emph{Second}, satellites are fast moving and experience larger propagation delays due to their greater physical distances. Here, novel channel models for \ac{THz} and \ac{mmWave} need to take into account these unique propagation environments as well as the considerable Doppler effects compared to terrestrial channel models. \emph{Third}, \ac{THz} links have pencil-beam directionality that necessitate the fine alignment of beams amid the increased Doppler and higher speeds in air and space. One approach that could facilitate the beam alignment process in a highly energy efficient fashion is the deployment of \acp{RIS} enabled \ac{THz} architectures, as discussed in  \cite{zhang2020millimeter} and \cite{tekbiyik2020reconfigurable}, and as further elaborated in Section IV-B. In particular, \acp{UAV} or satellites can carry a passive \ac{RIS} to improve air-to-air and air-ground links. Subsequently, based on the environment changes and the minute beam misalignments, the metasurfaces of these carried \acp{RIS} are continuously controlled to maintain 
reliable \ac{LoS} links. \emph{Fourth}, ubiquitous access through these systems necessitates the use of edge computing and storage. Nonetheless, computing problems will arise due to the on-board limitations of small satellites \cite{tataria20206g}. Thus, it necessary to develop novel networking frameworks that overcome the computing limitations and provide low latency \ac{THz} communications for \acp{NTN}.    
Other challenges that arise in \acp{NTN} include initial access, spectrum resource management, and cross-layer power management \cite{hu2020joint}. The design of such integrated space-air-ground systems is an open research area.
\section{Conclusion and Recommendations}
 In this paper, we have laid out a comprehensive roadmap outlining the seven defining features of \ac{THz} wireless systems that guarantee a successful deployment in next wireless generations. In particular, we have first presented a comprehensive overview of the fundamentals of \ac{THz}'s frequency bands. Based on these, we have examined the opportunities offered by the quasi-opticality of the \ac{THz}. Subsequently, we have investigated prospective architectures and essential network breakthroughs that can improve the connectivity of highly directional and \ac{LoS} dependent \ac{THz} links. Then, to guarantee a universal coverage and improve the network's scalability, we have scrutinized the synergy between the \ac{THz} band and lower frequency bands. Then, we have articulated the advantages, challenges, and opportunities surrounding joint sensing and communication systems. We have also investigated the techniques needed to guarantee a successful channel estimation and initial access process. Furthermore, we have also proposed novel \ac{ML} approaches to mitigate challenges surrounding network design and optimization. Then, we have shed light on specific auspicious \ac{THz} services that are expected to be the most anticipated technologies in the next decade. Clearly, \ac{THz} bands are still in their nascent stage and hold a promise for more revolutionary changes in the next-generation wireless networks.\\
 \indent Given the insights gathered from this comprehensive tutorial, we conclude with several \emph{recommendations} to achieve a successful \ac{THz} deployment and operation in next-generation wireless systems:
 \begin{itemize}
 	\item \textbf{Slow-Start:} Because of the short range links and the low accessibility to \ac{THz} \acp{SBS}, a first step towards the deployment of \ac{THz} wireless systems needs to occur in indoor areas or through the use of \acp{IS} overlaid on existing networks. 
 	\item \textbf{Towards Versatile Wireless Systems:} We envision that the success of future wireless \ac{THz} systems can be achieved by migrating towards versatile systems that have joint sensing and communication functions (and possibly other functions such as control and localization) as opposed to pure communication networks.
	\item \textbf{Towards Holographic Surfaces:} Investing in a massive amount of metasurfaces and avoiding cellular boundaries enables a richer sensing and more reliable communication and sensing links for \ac{THz} systems.
	\item \textbf{Prominent Role of Integrated Frequency Bands:} The inherent short communication range and intermittent links of \ac{THz} require a high-level of coexistence with sub-$\SI{6}{GHz}$ and \ac{mmWave} to deliver high-rate communication and high-resolution sensing in outdoor and mobile environments.
	\item \textbf{Pronounced Role of \ac{ML}:} Enabling successful channel estimation and a real-time network optimization at \ac{THz} systems requires novel out of the box \ac{ML} techniques that can address the peculiar properties of the \ac{THz} channel.
 \end{itemize}
\bibliographystyle{IEEEtran}
\bibliography{references}
\end{document}

%% file: acronyms.tex
\acro{D2D}{device-to-device}
\acro{SIR}{signal-to-interference-ratio}
\acro{SINR}{signal-to-interference-plus-noise-ratio}
\acro{PCP}{Poisson cluster process}
\acro{CoMP}{coordinated multi-point}
\acro{BS}{base station} 
\acro{MD-CoMP}{macrodiversity CoMP transmission}
\acro{MAC}{medium-access-control}
\acro{JT-CoMP}{joint transmission CoMP}
\acro{CoMP-JT}{coordinated multipoint joint transmission}
\acro{SBS}{small base station}
\acro{MDSD}{multiple devices to a single device}
\acro{MDS}{maximum distance separable}
\acro{SCN}{small cell network}
\acro{PPP}{Poisson point process}
\acro{TCP}{Thomas cluster process}
\acro{CSI}{channel state information}
\acro{PDF}{probability distribution function}
\acro{PMF}{probability mass function}
\acro{RV}{random variable}
\acro{i.i.d.}{independently and identically distributed}
\acro{MBMS}{multimedia broadcasting multicasting service}
\acro{EE}{energy efficiency}
\acro{HCP}{hard-core placement}
\acro{CCDF}{complementary cumulative distribution function}
\acro{CDF}{cumulative distribution function}
\acro{PC}{probabilistic caching}
\acro{RC}{random caching}
\acro{CPF}{caching popular files} 
\acro{PGFL}{probability generating functional}
\acro{KKT}{Karush-Kuhn-Tucker}
\acro{PGF}{point generating function}
\acro{SCA}{successive convex approximation}
\acro{HD}{high-definition}
\acro{FHD}{full-high-definition}
\acro{UHD}{ultra-high-definition}
\acro{VR}{virtual reality}
\acro{AR}{augmented reality}
\acro{5G}{fifth-generation}
\acro{QoS}{quality-of-service}
\acro{QoE}{quality-of-experience}
\acro{IoT}{Internet of Things}
\acro{MHCPP}{Matern hardcore point process}
\acro{LoS}{line-of-sight}
\acro{NLoS}{non-line-of-sight}
\acro{PSD}{power spectral density}
\acro{MEC}{mobile edge computing}
\acro{E2E}{end-to-end}
\acro{THz}{terahertz}
\acro{CLT}{central limit theorem}
\acro{HQ}{High Quality}
\acro{eMBB}{enhanced mobile broadband}
\acro{URLLC}{ultra reliable low latency communications}
\acro{mmWave}{millimeter wave}
\acro{EVT}{extreme value theory}
\acro{GEV}{generalized extreme value}
\acro{TVaR}{tail value at risk}
\acro{IoE}{Internet of Everything}
\acro{RL}{reinforcement learning}
\acro{GAN}{generative adversarial network}
\acro{HF}{high frequency}
\acro{RB}{resource block}
\acro{MIMO}{ multiple-input and multiple-output}
\acro{XR}{eXtended reality}
\acro{EM}{electromagnetic}
\acro{UE}{user equipment}
\acro{RIS}{reconfigurable intelligent surface}
\acro{IRS}{intelligent reflective surface}
\acro{HetNet}{heterogeneous network}
\acro{ML}{machine learning}
\acro{MLC}{machine learning for communications}
\acro{CML}{communications for machine learning}
\acro{RNN}{recurrent neural network}
\acro{MR}{mixed reality}
\acro{AoI}{age of information}
\acro{PAoI}{peak age of information}
\acro{RF}{radio frequency}
\acro{LCFS}{last come first served}
\acro{FCFS}{first come first served}
\acro{VaR}{value-at-risk}
\acro{CVaR}{conditional value-at-risk}
\acro{AVaR}{average value-at-risk}
\acro{ES}{expected shortfall}
\acro{QoPE}{quality of physical experience}
\acro{HRLLC}{high-rate and high-reliability low latency communications}
\acro{CRAS}{connected robotics and autonomous systems}
\acro{IoNT}{Internet of Nano-Things}
\acro{AoSA}{array of subarray}
\acro{LIS}{large intelligent surface}
\acro{IS}{information shower}
\acro{mMTC}{massive machine type communications}
\acro{OAM}{orbital angular momentum}
\acro{NOMA}{non-orthogonal mutliple access}
\acro{CPS}{cyber physical system}
\acro{UAV}{unmanned aerial vehicle}
\acro{DNN}{deep neural network}
\acro{VLC}{visible light communications}
\acro{OFDM}{orthogonal frequency division multiplexing}
\acro{OMA}{orthogonal multiple access}
\acro{LEO}{low earth orbit}
\acro{AI}{artificial intelligence}
\acro{TDD}{time division duplex}
\acro{CDMA}{code division multiple access}
\acro{OFDMA}{orthogonal frequency division multiple access}
\acro{NTN}{Non-Terrestrial Network}
\acro{SLAM}{simultaneous localization and mapping}